\documentclass{cta-author}

{}
{}
{}
\usepackage{multirow}
\usepackage{amsmath}
\usepackage{braket}
\usepackage{physics}
\usepackage{amssymb}
\usepackage{tabularx}
\usepackage{multicol}
\usepackage{graphicx}
\begin{document}
\supertitle{Submission Template for IET Research Journal Papers}

\title{Integrated semi-quantum layered communication}

\author{\au{Rajni Bala$^{\corr}$}, \au{Sooryansh Asthana}, \au{V. Ravishankar}}

\address{{Department of Physics, IIT Delhi, Hauz Khas, New Delhi,110016, India}
\email{Rajni.Bala@physics.iitd.ac.in}}
\begin{abstract}
In recent times, secure quantum communication in layered networks  has emerged as an 
important area of study. In this paper, we harness the potential offered by multidimensional states in secure quantum communication with only one quantum participant and all the other classical participants.
We propose three  protocols for--(i)  entanglement-based layered semi--quantum key distribution, (ii)  layered semi-quantum secret sharing, and, (iii)  integrated layered semi-quantum key distribution and secret sharing to share secret information in arbitrarily layered networks. These protocols integrate the features of semi–quantum communication in layered networks. All three protocols allow for {\it simultaneous} distribution of secure information in all the layers of a network, thanks to the employment of multidimensional states. We present these protocols for a small network of at most five participants and three layers and show the robustness of the same against various eavesdropping strategies. Finally, we provide a detailed
procedure for generalizations of the proposed protocols to distribute
keys/secrets in any arbitrarily structured quantum network. 
\end{abstract}
\maketitle
\section{Introduction}\label{introduction}

Since the proposal of the seminal BB84 protocol in 1984, the field of secure quantum communication has witnessed a great advancement \cite{Bennett84,Ekert91,pirandola2020advances}. Quantum key distribution (QKD) is the first application that has reached commercial levels (see, for example, \cite{RevModPhys.81.1301,liu2022towards} and references therein). This large research interest has resulted in several novel cryptographic protocols such as quantum secret sharing, and quantum secure direct communication that allow the distribution of secure information depending on realistic demands \cite{ hillery1999quantum,long2007quantum}.

 As a significant development, in 2007, a semi--QKD (SQKD) protocol \cite{Boyer07} has been proposed.  In this protocol, only one of the parties, {\it viz.}, Alice has access to quantum resources.  She can prepare any state and perform measurements in any basis. The other participant, {\it viz.}, Bob can measure only in the computational basis. For this reason, Bob is called a classical participant (CP).  The idea of SQKD is important-- both from theoretical and experimental viewpoints.  The theoretical interest stems from the question: what is the minimum quantumness needed to have a quantum advantage in secure communication protocols? From an experimental viewpoint, SQKD reduces the burden of the availability of quantum resources as only computational basis measurements are performed. That is, by employing already existing classical resources and minimum quantum resources, SQKD allows for implementing communication tasks.  These features make semi-quantum communication protocols appropriate for realizing in the current noisy-intermediate scale quantum (NISQ) regime \cite{yan2019semi,li2020new, massa2022experimental}.   These advantages have culminated not only into generalizations of SQKD to high-dimensional systems and multi-party systems but also into several semi-quantum communication protocols such as semi-quantum secret sharing (SQSS)  \cite{iqbal2020semi,li2013quantum,rong2021mediated,li2020new,yan2019semi,zhang17semiquantum,shukla2017semi}. 
  Quantum secret-sharing protocols have been designed for the distribution of secrets that are split between two participants. The protocol assumes that the individual behaviors of participants need not be honest, following \cite{hillery1999quantum}. For this reason, they should not be able to retrieve the information individually and can be retrieved only if the two collaborate.
 
 Elegant though these protocols are, the realistic situations involve multiple participants distributed in a quantum network. Recently, the study of the distribution of information securely over a network is a thriving field \cite{simon2017towards,epping2017multi,cho2021using}. In fact, quantum communication protocols such as QKD, quantum secret sharing, and quantum secure direct communication in networks have been proposed  \cite{pivoluska2018layered,qin2020hierarchical,qi202115,zhang2020tripartite}. In these protocols, all the participants are assumed to have access to quantum resources. However, to utilize existing resources in the NISQ era, it becomes worthwhile to distribute information securely with minimal quantum resources in networks. This approach has resulted in proposals of SQKD protocols in networks \cite{zhu2018semi, xian2009quantum}.

\begin{figure}[htb]
\centering
\includegraphics[width=0.25\textwidth]{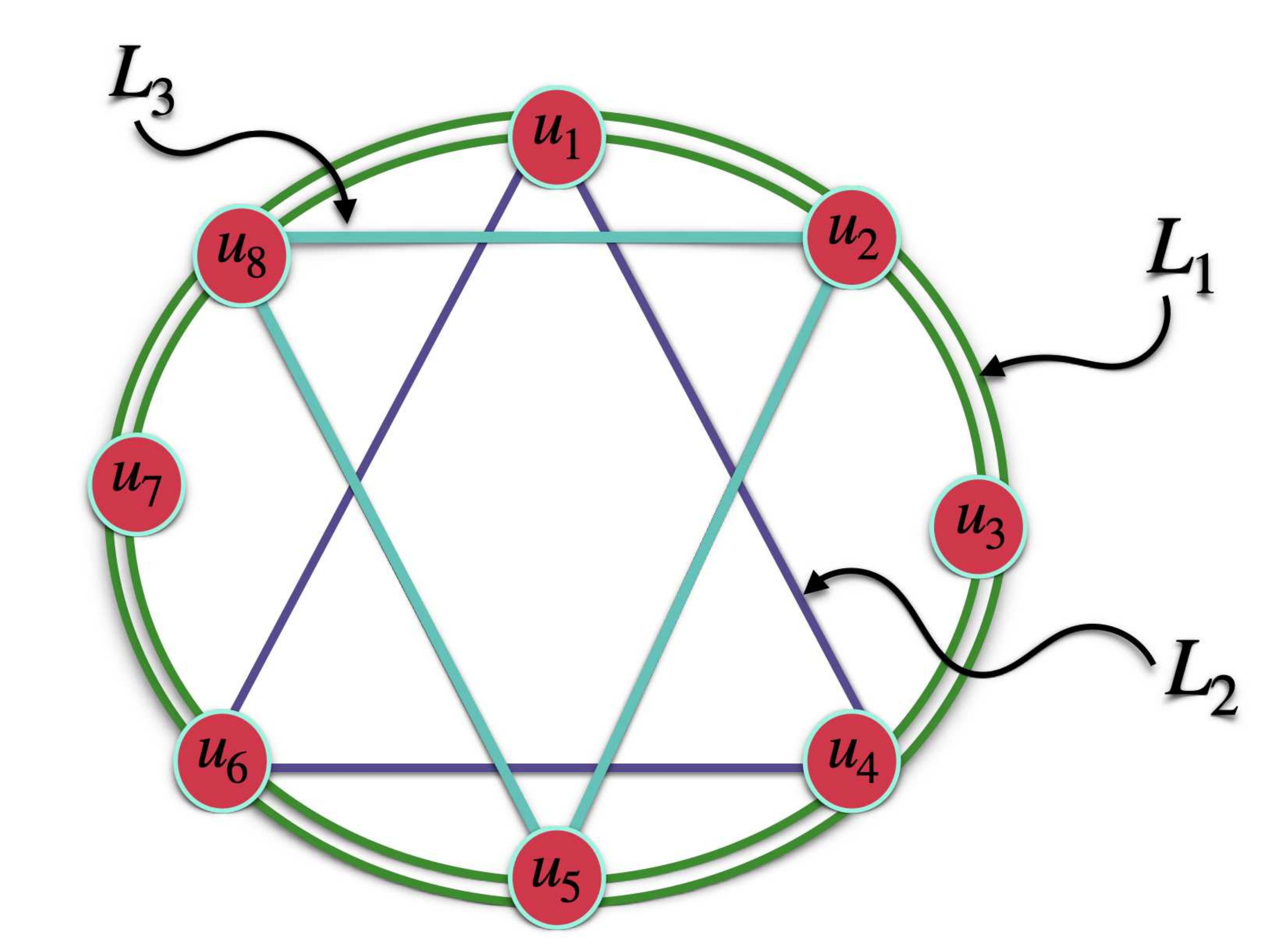}
\caption{Pictorial illustration of a layered quantum network: A network in which keys are to be shared in distinct layers $L_1, L_2$ and $L_3$. A participant may belong to more than one layer.  $L_1$  consists of participants $\{u_1, \cdots, u_8 \}$. Layers $L_2$ and $L_3$ consist of $\{u_1, u_4, u_6\}$ and $\{u_2, u_5, u_8\}$ respectively.}
\label{fig:layers}
\end{figure}

These developments and realistic demands have led us to focus on semi-quantum communication protocols in layered networks. By a layered network, we mean a network in which participants are divided into different groups, which we term `layers' (see figure (\ref{fig:layers}) for an example of a layered network). 
 Notably, these networks can be realized in various practical situations such as in banks. For example, suppose that a Bank manager needs to communicate with three subsets of employees- (i) with assistant managers, (ii) with probationary officers, and (iii) other communication with all the staff of the bank. These three form three layers of a network, \textit{viz.,} bank in this case. The academic sector provides us with another example of layered networks. Consider a situation in which a student has joined a doctoral program with two supervisors. The two supervisors need to communicate in private. Also, there may be a need for all three, i.e., the student and the two supervisors to communicate with one another. The former forms the 
layer $L_1$ and the latter forms the layer $L_2$. All such tasks require communication between different subsets of participants.

In this work, we show how  multidimensional states allow  \textit{simultaneous} distribution of secure information in all the layers. We have recently shown prepare-and-measure protocols for both--quantum and semi-quantum key distribution in layered networks \cite{bala2022quantum}. In this work, we take a step ahead and show how the appropriate choice of multidimensional states facilitates other task-dependent cryptographic communication. We illustrate it with three examples- (i) Entanglement-based layered SQKD (LSQKD), (ii) Layered SQSS (LSQSS), and (iii) Integrated layered semi-quantum key distribution and secret sharing (ILSKSS).

\begin{table}[h!]
 \begin{center}
 \resizebox{8cm}{3cm} {
\begin{tabular}{  |c| c| c| c| c|}
\hline
\multirow{3}{*}{}&\multirow{3}{3cm}{ \textbf{ Hierarchical secret sharing \cite{qin2020hierarchical}}}&\multirow{3}{4cm}{\bf Multiparty SQKD  \cite{tian2022multi} }&\multirow{3}{*}{{\bf Prepare-and-measure SQKD \cite{bala2022quantum}}}&\multirow{3}{2cm}{\bf Protocols proposed in this work} \\
 && &&\\
&&&&\\\hline
\multirow{3}{*}{ \textbf{Quantum channel}} &\multirow{3}{*}{\textbf{ideal}}&\multirow{3}{*}{\textbf{ ideal}}&\multirow{3}{*}{\textbf{ideal}}&\multirow{3}{*}{ \textbf{ideal}}\\
&&&&\\
&&&&\\\hline
\multirow{3}{2cm}{\textbf{ Resource states}} &\multirow{3}{2.5cm}{\textbf{multidimensional entangled states} }&\multirow{3}{3cm}{ \textbf{$2^{T+1} $ hyperentangled $~~~~~~~~~~$Bell states}}&\multirow{3}{3cm}{\textbf{multidimensional separable states}}&\multirow{3}{3cm}{\textbf{multidimensional entangled states}}\\
&&&&\\
&&&&\\ \hline
\multirow{3}{3cm}{\textbf{Network} }&\multirow{3}{3cm}{\textbf{ hierarchical}}&\multirow{3}{*}{{\textbf{ single layer} }}&\multirow{3}{*}{\textbf{multi-layered}}&\multirow{3}{3cm}{\centering\textbf{multi-layered}}\\
 &&&&\\
&&&&\\ \hline
\multirow{3}{3cm}{\textbf{ Quantum participant}}&\multirow{3}{3cm}{\centering\textbf{all}}&\multirow{3}{*}{\centering\textbf{one }}&\multirow{3}{*}{\centering\textbf{one}}&\multirow{3}{3cm}{\centering\textbf{ one}}\\
 &&&&\\
&&&&\\ \hline
\multirow{3}{3cm}{\textbf {Classical participants} }&\multirow{3}{3cm}{\centering\textbf  {none}}&\multirow{3}{*}{{\textbf  {$T$}}}&\multirow{3}{*}{\centering\textbf {all except one}}&\multirow{3}{3cm}{\centering\textbf {all except one}}\\
 &&&&\\
&&&&\\ \hline
\multirow{3}{3cm}{\textbf{ security against eavesdropping}}& \multirow{3}{*}{\centering\textbf{$\checkmark$ }}&\multirow{3}{3cm}{\centering\textbf{$\checkmark$}}& \multirow{3}{*}{\centering\textbf{$\checkmark$}}&\multirow{3}{*}{\centering\textbf{$\checkmark$}}\\
&&&&\\
&&&&\\\hline
\multirow{3}{3cm}{\textbf{Multiple tasks}}&\multirow{3}{*}{\centering\textbf{$\times$}}&\multirow{3}{*}{\centering\textbf{ $\times$}}&\multirow{3}{*}{\centering\textbf{$\times$}}&\multirow{3}{*}{\centering\textbf{$\checkmark$}}\\
&&&&\\ 
&&&&\\ \hline
\multirow{3}{3cm}{\textbf{Scope for generalized network }}&\multirow{3}{*}{\textbf{$\checkmark$}}&\multirow{3}{*}{\textbf{$\times$}}&\multirow{3}{*}{\textbf{$\checkmark$}}&\multirow{3}{*}{\textbf{$\checkmark$}}\\
&&&&\\
&&&&\\\hline
 \end{tabular}
 }
       \end{center}
    \caption{Comparison of the protocols proposed in this paper with the protocols\\ proposed in \cite{qin2020hierarchical,tian2022multi,bala2022quantum}.}
     \label{tab:Comparison_QKD}
\end{table}

The entanglement-based LSQKD protocol employs multidimensional entangled states to facilitate the secure distribution of keys in all the layers in one go (section (\ref{LSQKD})). LSQSS protocol employs multidimensional separable states of light to distribute secure information that can only be retrieved by collaboration among all the participants (section (\ref{LSQSS})). The third protocol, \textit{viz.}, ILSKSS is designed to have both features and employs multidimensional entangled states for distributing both keys and secrets simultaneously  (section (\ref{LSQSS+KD})). We first present these three protocols with examples of small networks and discuss the security of each against various eavesdropping attacks.  The resource states needed in the illustrative protocols have already been experimentally realized. For example, the three-photon asymmetric entangled states with Schmidt vectors  $(3,3,2)$ and $(4,4,2)$ have been prepared with fidelities of $80.1\%$ and $85.4\%$ in orbital angular momentum degree of freedom of photons respectively \cite{erhard2020advances,hu2020experimental}.

The generalizations of the protocols involve two steps- (a) identification of resource states, and (b) the steps of the protocol. We provide a methodology for the identification of resource states for an arbitrarily structured layered network (section (\ref{generalisation})). The steps of the protocol essentially remain the same as those of illustrative protocols. The major difference will be in the key/secret generation rules which we discuss separately (section (\ref{ksrule})). Though we present three protocols, the same procedure with appropriate modifications is amenable to other communication protocols depending on the situation at hand. In the table (\ref{tab:Comparison_QKD}), we compare various features of our protocols with already existing ones.  In \cite{pivoluska2018layered},  all the participants have been assumed to be quantum, and in \cite{qin2020hierarchical}, layered networks have not been considered. Additionally, the protocols proposed in \cite{pivoluska2018layered} and \cite{qin2020hierarchical} are not designed for integrated secure quantum communication. Rather they aim at only the secure distribution of keys. In this work, we propose protocols that are designed for integrated secure quantum communication in layered networks, that too with only one participant being quantum. In \cite{tian2022multi}, though the protocol allows for sharing a key with only one quantum participant, it does not allow sharing keys in different subsets of participants and so is not suitable for layered networks.   In the table (\ref{Protocols}), we show the type of resource states employed and the nature of participants in different protocols presented in the paper. For the sake of completeness, we very briefly discuss the concept of layered entanglement, quantum and classical participants, and SQKD protocol given by Boyer (\ref{preliminaries}).  Section (\ref{conclusion}) concludes the paper. 

 \begin{table}[h!]
 \begin{center}
 \resizebox{8.6cm}{2.2cm} {
\begin{tabular}{c  c c c c}
\hline
\multirow{5}{*}{{\bf S. No.}}&\multirow{5}{*}{\bf Protocol}&\multirow{5}{2cm}{\centering\bf State employed}&\multirow{5}{2.75cm}{\centering\bf Nature of participants (among whom keys/secrets are shared)} &\multirow{5}{1.85cm}{\bf Requirement of collaboration}\\
 &&&& \\
 &&&&\\
 &&&&\\
&&&&\\\hline
\multirow{3}{*}{1}&\multirow{3}{*}{\centering LSQKD} &\multirow{3}{*}{\centering entangled multiqudit}
&\multirow{3}{*}{\centering Honest}& \multirow{3}{*}{\centering No} \\
&&&&\\
&&&&\\\hline
 \multirow{3}{*}{\centering 2}& \multirow{3}{*}{\centering Layered--SQSS (LSQSS)}&\multirow{3}{*}{\centering separable multiqudit}&\multirow{3}{*}{\centering Dishonest}&\multirow{3}{*}{\centering Yes}\\
 &&&&\\
&&&&\\ \hline
 \multirow{5}{*}{\centering 3}& \multirow{5}{3cm}{\centering Integrated layered semi--quantum key distribution and secret sharing (ILSKSS)}&\multirow{5}{*}{\centering entangled multiqudit} &\multirow{5}{1.5cm}{\centering Honest and dishonest} & \multirow{5}{1.75cm}{\centering only among dishonest participants} \\
 &&&&\\
 &&&&\\
 &&&&\\
&&&&\\ \hline
 \end{tabular}
    \label{tab:general}
    }
    \end{center}
    \caption{Proposed protocols, nature of participants, and requirement of\\ collaboration.}
    \label{Protocols}
\end{table}
 
\section{Preliminaries}\label{preliminaries}
In this section, for the purpose of an uncluttered discussion, we briefly discuss quantum and classical participants and the  resources available to them. We next describe the SQKD protocol proposed by Boyer \cite{Boyer07}. Finally, we present layered entanglement \cite{malik2016multi} in pure states, which is a basic rudiment of all the protocols proposed in the paper, barring LSQSS. For a detailed discussion, please refer to \cite{malik2016multi,huber2013structure}.
 
\subsection{Quantum and classical paricipants}
\label{General setting}
\begin{enumerate}
    \item {\bf Quantum participant (QP):} A QP has full access to quantum resources. A QP is capable of preparing any state and performing any measurements and transformations. In the protocols, to reduce the burden of the availability of quantum resources, we consider only one QP named Alice. Given the task at hand, Alice plays a dual role, i.e., she may act as a participant of the network itself, or she may act as an external server who distributes the requisite states. 
    \item {\bf Classical participant (CP):} A CP has restricted capabilities. She  can prepare a state and measure it only in the computational basis. In the protocols, all the CPs are named Bob$_1$, Bob$_2$, $\cdots$, and so on.\\ 
        Following \cite{Boyer07}, classical participants can perform only two operations-- (i) CTRL, and (ii) Reflect.\\
    \noindent{\bf CTRL operation:} This corresponds to a classical participant performing a measurement in the computational basis, noting down the result, and sending the post-measurement state back to Alice.\\
    \noindent{\bf Reflect  operation:} In the Reflect operation, a classical participant simply resends the state to Alice.
\end{enumerate}

\subsection{Semi-quantum key distribution protocol proposed in \cite{Boyer07} (SQKD07 protocol): a brief recap}
SQKD protocol, proposed by Boyer \textit{et al.}, is designed for the distribution of  a secure key between two participants, {\it viz.},  Alice and Bob. Please note that Bob is a classical participant.
Since the SQKD protocol has only one QP, it eases down the experimental implementation of the SQKD07 protocol \cite{massa2022experimental}. The steps of the SQKD07 protocol are explicitly enumerated as follows:
\begin{enumerate}
    \item Alice randomly prepares one of the states from the two basis sets $\{\ket{0}, \ket{1}\},~ {\rm and}~\{\ket{+}, \ket{-}\}$ and sends it to Bob. 
    \item Bob either performs (a) CTRL operation (measures the received state in the computational basis and sends the post-measurement state)  or  (b) { Reflect operation} (simply sends the received state back to Alice). 
    \item Alice measures the received state in the same basis in which it was prepared initially. This constitutes one round and this process is repeated for a sufficiently large number of rounds.
    \item  On a classically authenticated channel, Bob reveals the rounds in which he has measured and Alice reveals the rounds in which she has sent the states in the computational basis, i.e., $\{\ket{0}, \ket{1}\}$. 
    \item Alice analyses the data of those rounds in which Bob has not measured to check for the presence of an eavesdropper.
    \item In the absence of an eavesdropper, the outcomes of the rounds, in which Bob has measured and Alice chooses the computational basis, constitute a key.
\end{enumerate}
In this way, Alice and Bob shared a secure key.

\subsection{Layered entanglement}
  Multipartite high-dimensional entangled states have a rich structure of entanglement. There may exist multiparty high-dimensional entangled states in which different parties do not necessarily have subsystems belonging to identical dimensions. This property has been termed layered entanglement.  In these states, different subsystems belong to Hilbert spaces of different dimensions. This  variation in the dimensionalities  of subsystems owes to the asymmetric nature of entanglement amongst different parties  which is termed as layered entanglement \cite{huber2013structure}.
  For example, consider the state,
\begin{align}
    |\Psi\rangle\equiv \frac{1}{2}\Big(|000\rangle+|111\rangle+|220\rangle+|331\rangle\Big).
    \label{eqn:QKDstate}
\end{align}
The Schmidt ranks for the first, second, and third subsystems are 4, 4, 2 which may be  arranged as a vector, $(4,4,2)$, known as {\it Schmidt vector} of the state $|\Psi\rangle$. The nature of correlation possessed by the state can be understood as follows:\\
The first two subsystems have a perfect correlation with each other and also with the third subsystem in the computational basis. That is to say, if a measurement is performed on the first subsystem in the computational basis, the outcome uniquely specifies the outcomes in the measurement basis  of the rest of the two subsystems. However, this is not the case with the third subsystem. This particular feature owes to asymmetric entanglement in different subsystems and is termed layered entanglement.  

Equipped with this information, we now move on to present three different communication protocols. We start with SQKD in layered networks.
\section{Entanglement-based layered semi--quantum key distribution (LSQKD)}
\label{LSQKD}

\begin{figure}[h!]
\centering
\includegraphics[width=0.3\textwidth]{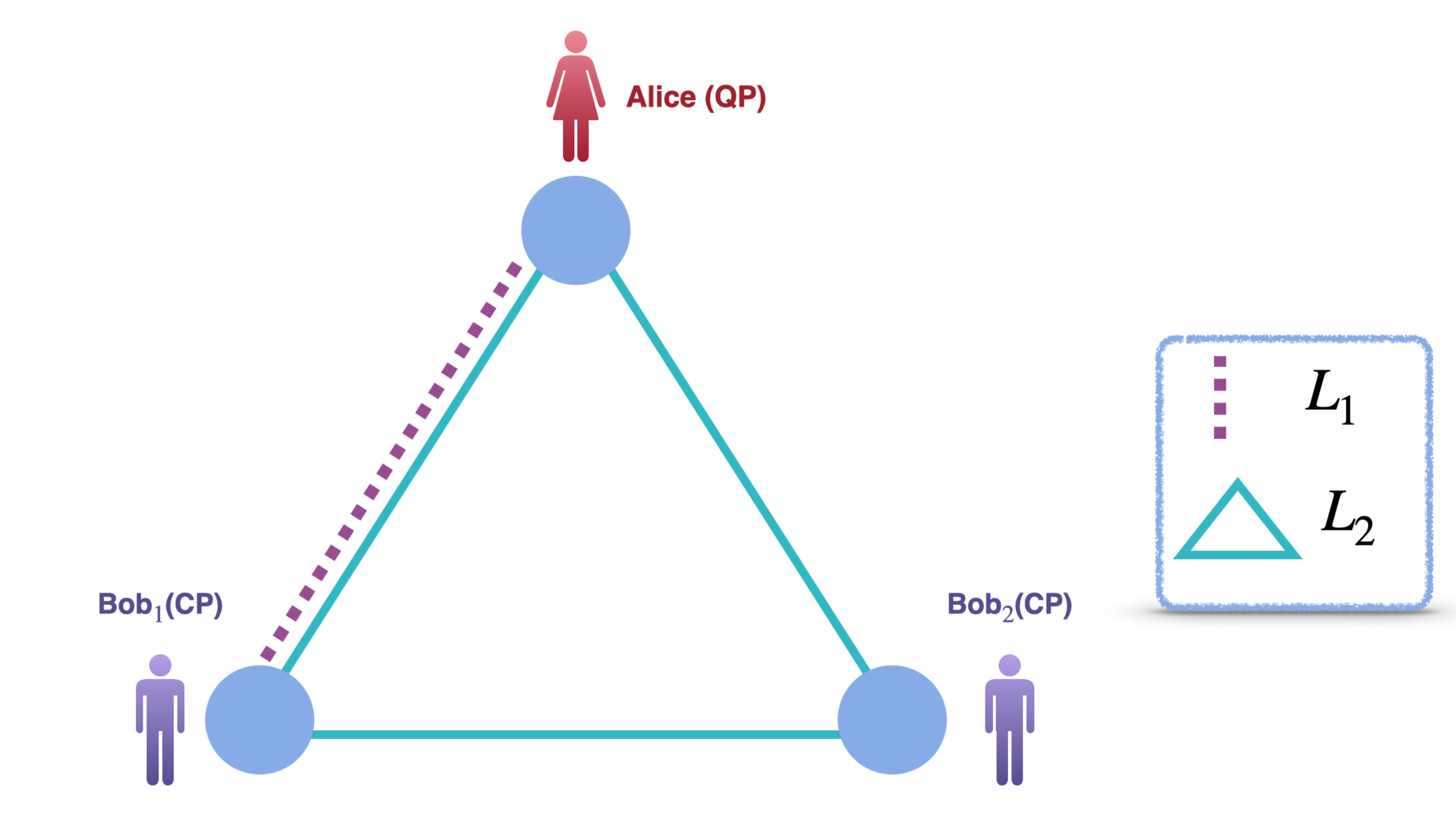}
\caption{A network of three participants having two layers $L_1$ and $L_2$. The layer $L_1$ consists of Alice and Bob$_1$ (shown by a dotted line).  The layer $L_2$ consists of all the three participants, {\it viz.}, Alice, Bob$_1$, and Bob$_2$ (shown by a triangle).}
\label{fig:QKD}
\end{figure}

In this section, we present an entanglement-based LSQKD protocol for a network of three participants Alice (QP),  Bob$_1$ (CP), and Bob$_2$ (CP), and two layers $L_1$ and $L_2$. Layer $L_1$ consists of two participants, {\it viz.},  Alice, and  Bob$_1$, while layer $L_2$ has all three participants. The description of the network is pictorially shown in the figure (\ref{fig:QKD}).

\noindent\textbf{\textit{Aim:}} Simultaneous distribution of keys in both layers $L_1$ and $L_2$. 

\noindent{\textbf{\textit{Resources:}}} Tripartite $(4\otimes4\otimes 2)$ dimensional entangled state (same as the one given in equation (\ref{eqn:QKDstate})),
\begin{equation*}
|\Psi\rangle\equiv\frac{1}{2}\Big(|000\rangle+|111\rangle+|220\rangle+|331\rangle\Big), 
\end{equation*}
which is shared among Alice, Bob$_1$, and Bob$_2$. The first two subsystems are four-level systems and the third subsystem is a two-level system.  The rationale underlying this choice of the state is explained in Appendix (A1).

\begin{figure}[h!]
\includegraphics[width=0.45\textwidth]{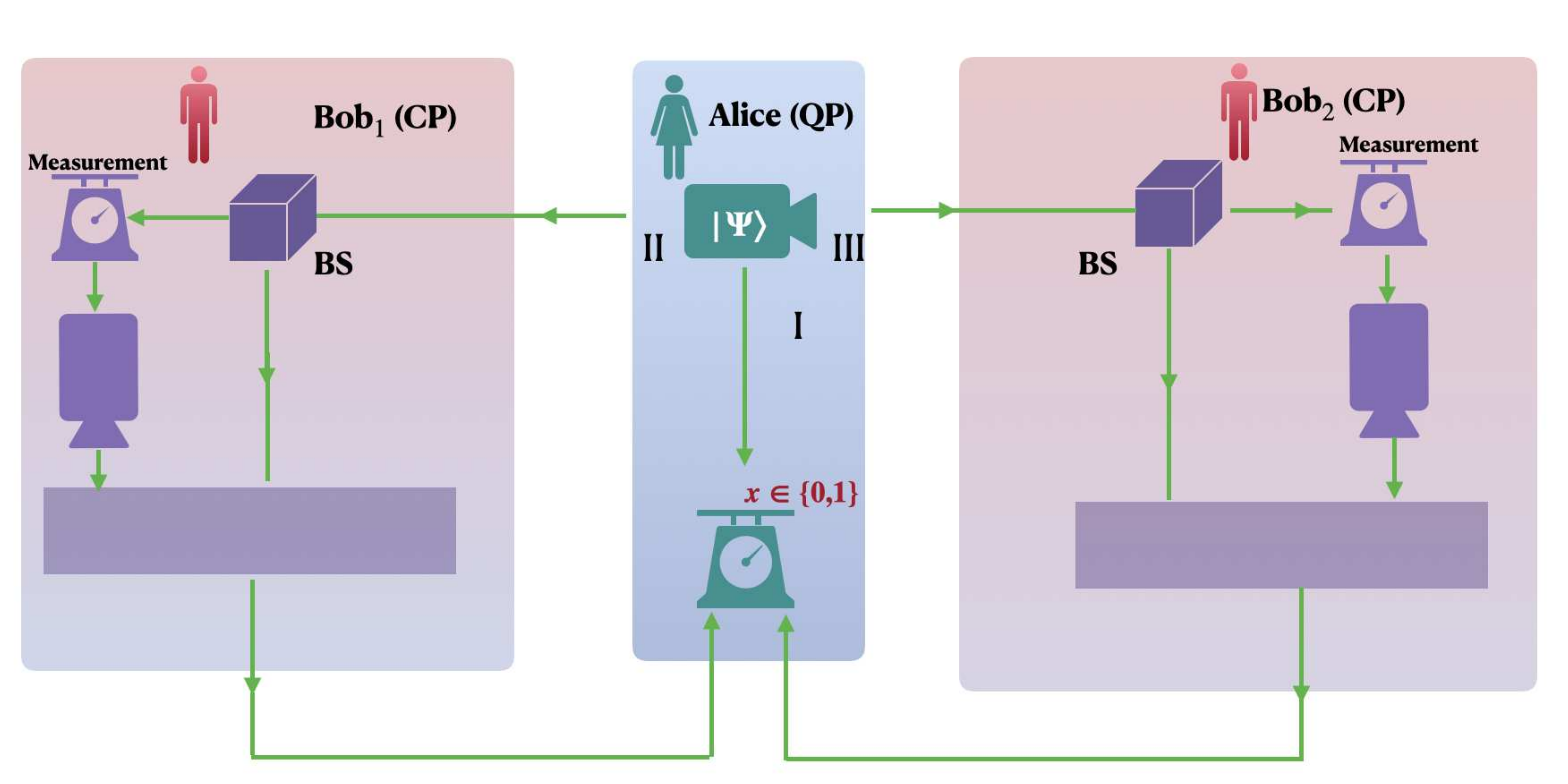}
\captionof{figure}{Pictorial representation of LSQKD protocol reflecting actions of each participant. Alice (QP) prepares the state $\ket{\Psi}$, keeps the first subsystem with herself and sends its second and third subsystems to Bob$_1$ and Bob$_2$ respectively. Bob$_1$ and Bob$_2$ either perform a measurement in the computational basis or send the incoming state back with equal probability. Alice performs a measurement of $\ket{\Psi}\bra{\Psi}$ (corresponding to $x=0$) or performs a measurement in the computational basis (corresponding to $x=1$).}
\label{fig:Steps} 
\end{figure}

\begin{center}
    \textbf{ The protocol}
\end{center}
The steps of the protocol are as follows:
\begin{enumerate}
    \item Alice, being QP,  prepares the state $|\Psi\rangle$ and sends the second and the third subsystem to CPs Bob$_1$ and Bob$_2$ respectively. 
    \item Bob$_1$ and Bob$_2$, upon receiving the states, randomly choose the following two actions with equal probability:\\
    (a) \textbf{CTRL:} This corresponds to measuring in the computational basis and sending the post-measurement state to Alice,\\  (b) \textbf{Reflect:} This corresponds to simply sending the received state to Alice.
    \item Alice, upon receiving the subsystems from both the Bobs, either performs measurements on the three subsystems in the computational basis or performs a projective measurement of $\Pi_{\Psi}\equiv|\Psi\rangle\langle\Psi|$. 
    \item This completes one round. This process is repeated for many rounds.
    \item After a  sufficient number of rounds, Bob$_1$ and Bob$_2$ reveal the rounds  in which they have performed CTRL operation and Alice reveals the rounds in which she has performed measurements in the computational basis.
    \item Alice analyses the data of those rounds in which none of the Bobs have performed CTRL operations to check for the presence of an eavesdropper.
    \item In the absence of any eavesdropper, the rounds, in which all the Bobs of a layer have performed CTRL operation and Alice has measured in the computational basis, constitute a key in the respective layer.
    \end{enumerate}
The schematic diagram of the protocol is shown in the figure (\ref{fig:Steps}).

\subsection{Key generation rule} \label{key_rule}
To obtain a key symbol in each layer, Alice, Bob$_1$, and Bob$_2$ employ binary representations of their measurement outcomes. Let   $a,~b_1,~b_2$ be  the outcomes of Alice, Bob$_1$, and Bob$_2$ respectively. In the binary representation\footnote{The outcome $b_2$ of Bob$_2$ is already in binary form.}, these are expressed as:
\begin{equation}\label{eqn:kgrule}
    a=2a^{(1)}+a^{(0)},\quad b_1=2b_1^{(1)}+b_1^{(0)},
\end{equation}
 where symbols $a^{(0)}$ and $b_1^{(0)}$ are coefficients of $2^0$ and symbols $a^{(1)}$ and $b_1^{(1)}$ are coefficients of $2^1$. 
The symbols $a^{(1)},~b_1^{(1)}$ constitute a key in  layer $L_1$ whereas the symbols $a^{(0)},~b_1^{(0)},~b_2$, in layer $L_2$.\\
Since two key symbols are generated with a uniform probability, the sifted key rate in both layers is $1$ bit.

\subsection{Confidentiality of the key}
Following the above prescription, keys in both layers can be shared simultaneously. However, it is important that the two keys remain confidential. We argue that the two keys are indeed confidential, i.e., any outsider participant who does not belong to a particular layer cannot retrieve any information about the key being shared. 

So, for the protocol, it suffices to ensure that Bob$_2$ cannot gain any information about the key being shared in layer $L_1$.
 Equation (\ref{eqn:QKDstate}) clearly indicates that for each outcome of Bob$_2$, there are two outcomes for Alice and Bob$_1$. These two outcomes correspond to two different key symbols following equation (\ref{eqn:kgrule}). For example, for an outcome $0$ at Bob$_2$'end, Alice and Bob$_1$ obtain two outcomes $0$ and $2$ with equal probabilities. These outcomes correspond to key symbols $0$ and $1$ in layer $L_1$ with uniform probability. A similar analysis holds for the case when Bob$_2$ gets outcome $1$. Thus, Bob$_2$ gains no information on either Alice or Bob$_1$'s outcomes and hence on keys being shared in layer $L_1$. Thus, keys are shared confidentially.

  \subsection{Security against various attacks}
 \label{security}
 In this section, we analyze the security of LSQKD against various eavesdropping strategies.
Since the protocol is designed for the simultaneous distribution of keys in both layers, Eve may be interested in knowing the key being shared either in one of the layers or in both layers. The former kind of eavesdropping attack is similar to the one encountered in single-layer SQKD protocols. The latter, however, is unique to the structure of the network and hence deserves a detailed discussion.
Since Eve wishes to obtain maximum information with minimal errors, she would like to identify the participant who belongs to both layers instead of separately attacking both layers. In the LSQKD protocol, Bob$_1$ is the participant who belongs to both layers, so Eve would try to attack the subsystem traveling to him. To start with, we consider an intercept-resend attack.\\

 \subsubsection{Intercept-resend attack} In this attack, Eve intercepts the states traveling to Alice and measures in the computational basis. Afterward, she sends the post-measurement state to Bob$_1$. Since Bob$_1$ is a CP, he either performs CTRL or  Reflect operations. The rounds in which Bob$_1$ performs CTRL operation, Eve obtains full information being shared in both layers. However, in half of the rounds, Bob$_1$ performs Reflect operation, which helps detect Eve's interventions. This is because Alice, upon receiving states from both Bobs, measures either in the computational basis or performs projective measurements on $\ket{\Psi}\bra{\Psi}$. So, due to Eve's interventions, the state gets disturbed and Alice will not get the desired result when she performs a  projective measurement $\ket{\Psi}\bra{\Psi}$. So, whenever both Bobs perform Reflect operation and Alice performs a projective measurement $\ket{\Psi}\bra{\Psi}$, Eve's interventions introduce errors with a probability of $0.75$. Therefore, in $l $ such rounds, Eve gets detected with a probability of $1-(0.25)^l$.

 However, if Eve attacks Bob$_2$'s subsystems, her interventions get detected with a probability of $0.5$. For $l$ such rounds, Eve gets detected with a probability $1-(0.5)^l$.

\subsubsection{Entangle-and-measure attack} Eve's entangle-and-measure strategy has the same impact on the protocol as that of the intercept-resend strategy. If Eve entangles the subsystem traveling to Bob$_1$, its action in the computational basis can be expressed as:
\begin{align}
    \ket{\Psi}\ket{0}_E\xrightarrow[]{U} \frac{1}{2}&\Big(\ket{000}\ket{0}_E+\ket{111}\ket{1}_E\nonumber\\
+&\ket{220}\ket{2}_E+\ket{331}\ket{3}_E\Big).
\end{align}
The above equation shows that Eve gains the same information as Bob$_1$. However, her presence gets detected in those rounds in which both Bobs perform Reflect operation and Alice performs a projective measurement $\ket{\Psi}\bra{\Psi}$. So, Eve gets detected with a probability of $0.75$. For $l$ such rounds, the probability of Eve's detection is $(1-0.25^l)$ which approaches unity for a sufficiently large $l$.

In the above analysis, we have considered those attacks in which Eve employs attack when states are traveling from Alice to respective Bobs. However, semi-quantum protocols are two-way communication protocols. That is to say, the subsystems travel from Alice to respective Bobs and vice-versa. This feature provides Eve with two opportunities to interact with the systems. Therefore, in the following, we consider a two-way general entangling attack.\\

 \subsubsection{Two-way entangling attack}
 In this attack, Eve interacts her ancillae with traveling subsystems. She may employ two different unitaries to interact with  subsystems traveling to and from the respective Bob.  

\begin{figure}[h!]
\centering
\includegraphics[width=0.45\textwidth]{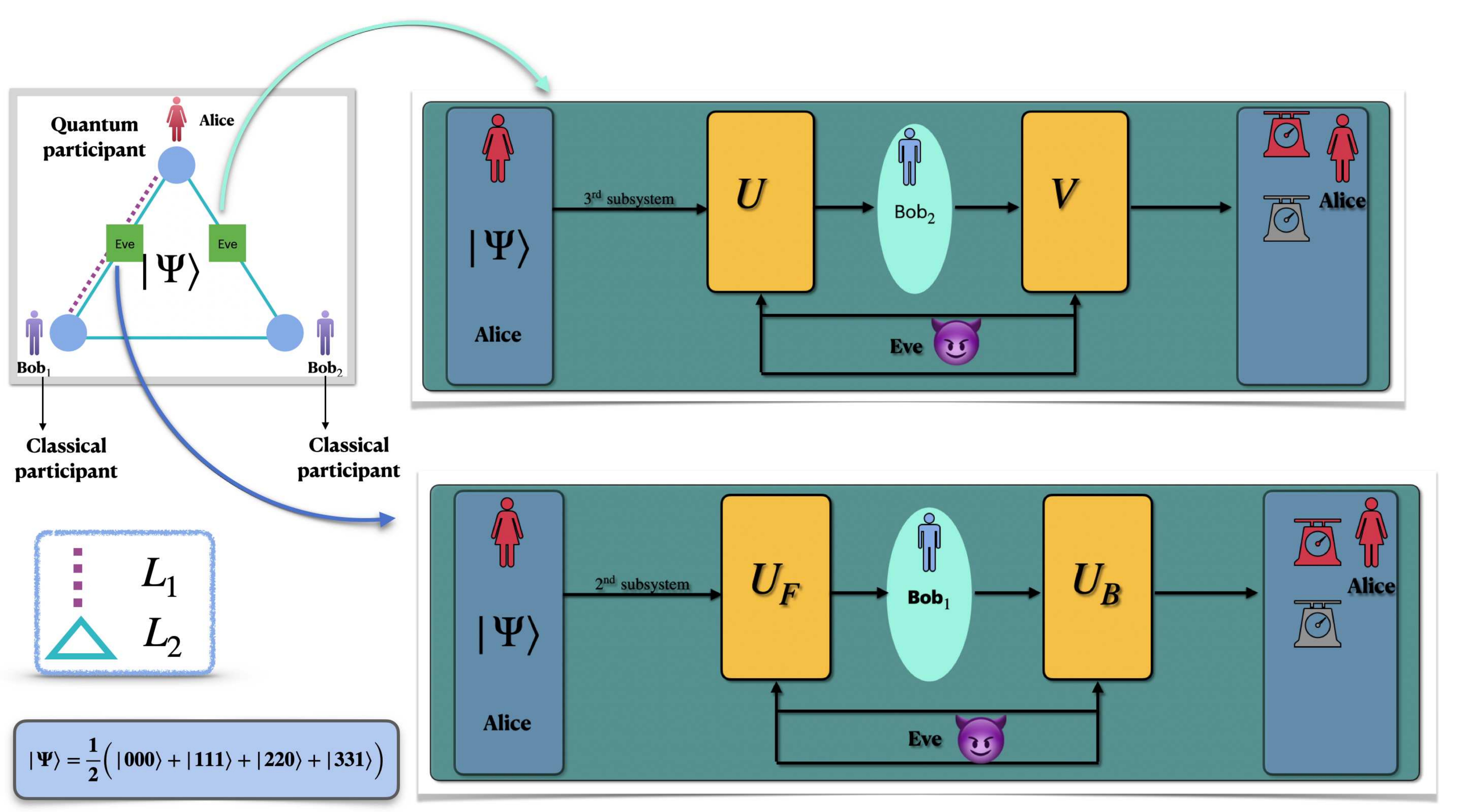}
\caption{Pictorial representation of  a two-way entangling eavesdropping strategy on a layered network. Eve may attack either on the state traveling to Bob$_1$ or Bob$_2$, that too either in the forward path or in the backward path.  }\label{fig:two-way}
\end{figure}

As discussed earlier, Eve may attempt to gain information being shared in only one of the layers or in both layers.  A pictorial representation of these two strategies  is shown in figure (\ref{fig:two-way}). The case in which Eve wishes to gain information being shared in both layers, it will be optimal for her to attack Bob$_1$. We discuss both cases as follows:\\

\noindent{\textbf{\textit{ Case I: Attack in layer $\boldsymbol{L_2}$:}}} In this case, it suffices for Alice to attack the subsystem traveling to and from Bob$_2$. Suppose that Eve employs the unitary $U$ for the subsystem traveling to Bob$_2$ and $V$ for subsystems traveling from Bob$_2$.
So, the combined state after Eve's interaction $U$ can be expressed as:
 \begin{align}
     \ket{\Psi}\ket{0}_ E\xrightarrow[]{U}\frac{1}{2}&\Big\{\big(\ket{00}+\ket{22}\big)\ket{0}\ket{E_{0}}\nonumber\\
     +&\big(\ket{11}+\ket{33}\big)\ket{1}\ket{E_{1}}\Big\},
 \end{align}
 where $|E_{i}\rangle$ are the  states of Eve's ancilla after interaction satisfying conditions imposed by unitary $U$. 
 The action of Eve's second interaction $V$ in the computational basis  can be expressed as:
\begin{align} \label{eq:backBob2}  \ket{i}\ket{0}_{E_2}\xrightarrow[]{V}\sum_{j=0,1}\ket{j}\ket{F_{ij}}, ~~~~i\in\{0,1\},
\end{align}
where $\ket{F_{ij}}$ are states of Eve's ancilla after interaction with unitary $V$.
 
 Since the participants perform CTRL and Reflect operations randomly, we discuss the two cases one by one as follows: \\

\noindent{\it (a)  Bob$_2$ performs CTRL operation:} 
When Bob$_2$ performs a CTRL operation, his post-measurement state is in the computational basis which he sends to Alice. Alice can detect this tampering when she measures in the computational basis. For Bob$_2$'s post-measurement state $\ket{i}$, Eve gets  detected with a probability of $\norm{\ket{F_{i\bar{i}}}}^2$ (${\bar i} = i+1 ~{\rm mod}~2$). \\

 \noindent{\textit{(b) Bob$_2$ performs Reflect operation:}}
 This corresponds to the rounds, when both Bobs perform Reflect operation, i.e.,  they send the received state back to Alice. In that case, the action of Eve's tampering on the combined state can be expressed as:
 \begin{align}
     \ket{\Psi}\ket{0}_{E}\ket{0}_F\xrightarrow[]{VU}&\frac{1}{2}\Big\{\Big(\ket{00}+\ket{22}\Big)\Big(\sum_{j=0,1}\ket{j}\ket{E_0}\ket{F_{0j}}\Big)\nonumber\\&+\Big(\ket{11}+\ket{33}\Big)\Big(\sum_{j=0,1}\ket{j}\ket{E_1}\ket{F_{1j}}\Big)\Big\}
 \end{align}

If Alice measures in the computational basis, Eve's tampering gets detected with a probability of $\frac{1}{2}\Big( \norm{\ket{E_0}\ket{F_{01}}}^2+\norm{\ket{E_1}\ket{F_{10}}}^2\Big)$.
If Alice performs a projective measurement $\ket{\Psi}\bra{\Psi}$, she gets the correct result with a probability, $p=\frac{1}{4}\norm{\ket{E_0}\ket{F_{00}}+\ket{E_1}\ket{F_{11}}}^2$. Thus, Eve gets detected with a probability $(1-p)$.
 
 A similar analysis holds if Eve attacks only on layer $L_1$. We now discuss the case when Eve simultaneously attacks on both layers. In this case, it suffices for Eve to attack on the subsystem traveling to and from Bob$_1$. This is because Bob$_1$ belongs to both layers, so gaining his information is equivalent to gaining information being shared in both layers.\\

 \noindent{\textbf{\textit{ Case II: Simultaneous attack on both layers $\boldsymbol{L_1}$ and $\boldsymbol{L_2}$:}} Suppose that Eve employs the unitary $U_F$ on the subsystem traveling to Bob$_1$ and the unitary $U_B$ on the subsystem traveling from Bob$_1$.
 Suppose that the combined state shared among participants and Eve, after unitary $U_F$, can be expressed as:
 \begin{align}
     \ket{\Psi}\ket{0}_F\xrightarrow[]{U_F} \frac{1}{2}&\Big\{\ket{000}\ket{E_0}+\ket{111}\ket{E_1}\nonumber\\
     +&\ket{220}\ket{E_2}+\ket{331}\ket{E_3}\Big\},
 \end{align}
 where $\ket{E_i}$ are states of Eve's ancilla after interaction satisfying constraints imposed by unitary $U_F$. Since Bob$_1$ either performs CTRL or Reflect operation, we discuss the two cases separately. Before doing that, suppose that the action of unitary $U_B$ in the computational basis can be expressed as:
 \begin{align}\label{eq:backword}
     \ket{i}\ket{0}_B\xrightarrow[]{U_B}\sum_{j=0}^3\ket{j}\ket{F_{ij}}, ~~~~~~~~i\in\{0,1,2,3\}.
 \end{align}
The state $\ket{0}_B$ is the state of Eve's ancilla before interaction and  $\ket{F_{ij}}$ represents an unnormalized and non-orthogonal state of Eve's ancilla after the interaction $U_B$. \\

\noindent{\textit{(a) Both Bobs perform CTRL operation:}} 
When Bob$_1$ performs CTRL operation, he gets the post-measurement state $\ket{i} $, where $i\in\{0,1,2,3\}$ with a probability of $0.25$. He sends this state to Alice which, due to Eve's tampering changes following the equation (\ref{eq:backword}). This tampering of Eve gets detected when Alice measures in computational basis with a probability of $\sum_{j\neq i}\norm{\ket{F_{ij}}}^2$.\\

\noindent{\textit{(b) Bob$_1$ performs Reflect operation:}}
When both Bobs perform Reflect operation, the combined state of all the participants and Eve after Eve's two unitary interactions $U_F$ and $U_B$ can be expressed as:
\begin{align}    &\ket{\Psi}_{ab_1b_2}\ket{00}_{EF}\xrightarrow[]{U_BU_F}\frac{1}{2}\Big\{\sum_{j=0}^3\Big(\ket{0}_a\ket{j}_{b_1}\ket{0}_{b_2}\ket{E_0}\ket{F_{0j}}\nonumber\\
&~~~+\ket{1}_a\ket{j}_{b_1}\ket{1}_{b_2}\ket{E_1}\ket{F_{1j}}+\ket{2}_a\ket{j}_{b_1}\ket{0}_{b_2}\ket{E_2}\ket{F_{2j}}\nonumber\\
&~~~~+\ket{3}_a\ket{j}_{b_1}\ket{1}_{b_2}\ket{E_3}\ket{F_{3j}}\Big)\Big\}
\end{align}

In the above equation,  the subscripts $a$, $b_1$, $b_2$, $E$, and $F$ represent subsystems belonging to Alice, Bob$_1$, Bob$_2$, Eve's ancillae in forward, and backward paths respectively and are used to avoid confusion.\\
In this case, whenever Alice measures in the computational basis and her post-measurement state is $\ket{i}$, Eve's interventions get detected with a probability of $    \frac{1}{4}\sum_{i,j=0,\newline j\neq i}^3\norm{\ket{E_i}\ket{F_{ij}}}^2$.

However, the rounds in which Alice performs a  projective measurement $\ket{\Psi}\bra{\Psi}$, she gets the correct result with a probability, $p=\frac{1}{16}\norm{\sum_{i=0}^3\ket{E_i}\ket{F_{ii}}}^2$. Thus, Eve's actions get detected with probability $(1-p)$.

 In this way, different interventions of Eve will get detected and hence, the proposed protocols are secure against such attacks.


 We conclude this section with the following remark. The employment of layered entangled states and  semi--quantum secure communication protocols are not exclusive to key distribution. In fact, the  proposals of \cite{qin2020hierarchical,rong2021mediated,li2013quantum,sun2009multiparty,li2020new,yan2019semi} testify to the generality of the secure communication with classical Bobs. {\it Layered} secure semi--quantum communication protocols, however, remain much less explored. In what follows,  we make a foray into these protocols one by one.


\section{Layered semi--quantum secret sharing (LSQSS)}\label{LSQSS}
In the previous section, we show that by employing multidimensional entangled states, keys can be distributed securely in multiple layers of a network having only one QP. Such protocols are useful when the individual behavior of all the participants is honest. We now move on to another situation, first introduced by Hillary, for two participants \cite{hillery1999quantum} in which the individual behavior of participants may not be honest. For this reason, the  secrets 
should not be retrieved unless all the participants collaborate. In the following, we show that using multidimensional separable states these secrets can be distributed in all the layers of a network in one go.\\

\noindent{\textbf{\textit{Network:}}} A network of five participants, {\it viz.}, Bob$_1$, Bob$_2$, Bob$_3$, Bob$_4$ and Bob$_5$ and three layers. Layer $L_1$ consists of Bob$_1$ and Bob$_2$. Layer $L_2$ consists of Bob$_3$, Bob$_4$, and Bob$_5$. The third layer \textit{viz.,} $L_3$ consists of all five participants. Alice is the external QP who is given the task to distribute three random secrets. The pictorial representation of the network is given in figure (\ref{fig:qss}).\\ 

\begin{figure}[h!]
\centering
\includegraphics[width=0.4\textwidth]{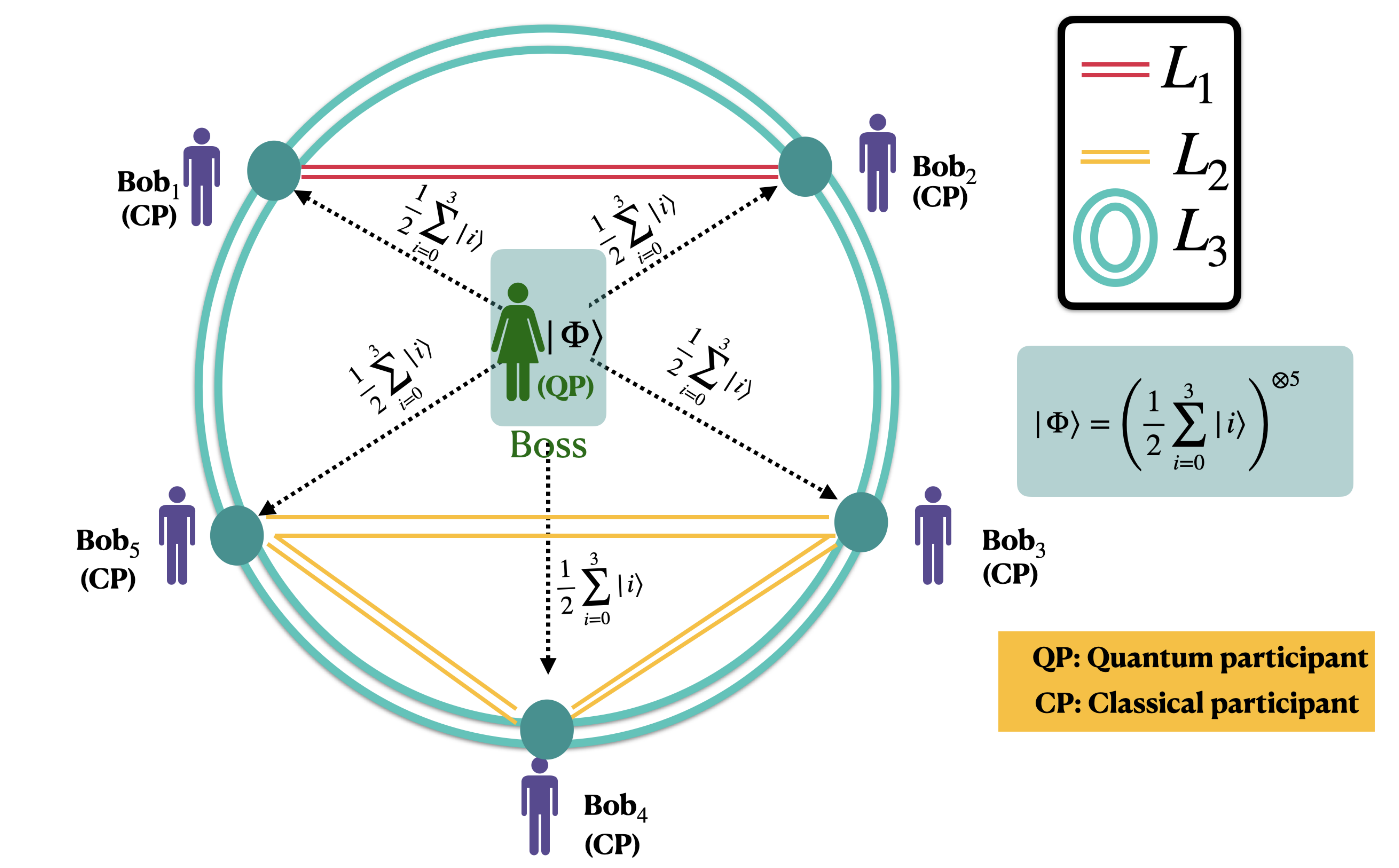}
\caption{Pictorial representation of a network having five CPs and three layers. Different layers are shown in the inset.}
\label{fig:qss}
\end{figure}

\noindent{\textbf{\textit{Aim:}}} Simultaneous distribution of secrets in three layers in a fully confidential manner.\\
\noindent{\textbf{\textit{Resources:}}} $4\otimes 4\otimes 4\otimes 4\otimes4$ dimensional separable state, 
\begin{align}\label{eq:SQSS}
    |\Phi\rangle\equiv &~~\Big(\frac{1}{2}({|0\rangle+|1\rangle+|2\rangle+|3\rangle})\Big)^{\otimes 5}.
\end{align}
The rationale for this choice of state has been given in Appendix (A2). The resource state used is a separable state, which can be prepared with a spatial light modulator (SLM) in OAM degrees of freedom. 
\begin{center}
    \textbf{The protocol}
\end{center}
The steps of the protocol are as follows:
\begin{enumerate}
      \item Alice, being QP, prepares the state $|\Phi\rangle$ and sends the $i^{{\rm th}}$ subsystem to  Bob$_i,~i \in \{1, \cdots, 5\}$.
    \item Each Bob, on receiving the state, performs either CTRL or Reflect operation randomly.\\ Please recall that \textit{CTRL} operation is to measure in the computational basis and send the post-measurement state to Alice, while  \textit{Reflect} operation is to simply send the received state to Alice.
      \item Alice, upon receiving the state from each Bob, either performs a measurement on each subsystem in the computational basis or performs a projective measurement $\Pi_\Phi \equiv |\Phi\rangle\langle\Phi|$. 
      \item Steps $(1-3)$ constitute one round and are repeated for a sufficiently large number of rounds. 
    \item Afterwards, each Bob reveals the rounds in which CTRL operations are performed. Alice reveals the rounds in which she has measured in the computational basis.
    \item Alice analyses the data of those rounds in which none of the Bobs have performed CTRL operations to check for the presence of  eavesdropping.
    \item In the absence of  eavesdropping, the outcomes of the rounds in which all the Bobs of a given layer have performed CTRL operations are utilized to generate a secret. As required, the participants can retrieve this secret only if they collaborate among themselves.
   \end{enumerate}

   \subsection{Secret generation rule}
 
 As in the key generation rule of the LSQKD, each participant  writes his measurement outcomes in binary representation. Let outcomes of Bob$_i$ be $b_i$ ($1\leq i \leq 5$). In the binary representation, these are expressed as:
 \begin{align}
 b_i\equiv 2b_i^{(1)}+b_i^{(0)}, \quad\quad\quad\quad ~~~~~~ i \in \{1, \cdots, 5\}.
 \end{align}
  The secret symbols in layers $L_1$ and $L_2$ are defined as $s_1\equiv b_1^{(1)}\oplus b_2^{(1)}$ and $s_2\equiv b_3^{(1)}\oplus b_4^{(1)}\oplus b_5^{(1)} $ respectively, where the symbol $\oplus$ represents the operation of addition modulo 2. Similarly, a secret in layer $L_3$ having all the five participants is defined as, $s_3\equiv b_1^{(0)} \oplus b_2^{(0)}\oplus b_3^{(0)}\oplus b_4^{(0)}\oplus b_5^{(0)}$.

Similar to LSQKD, secrets shared in LSQSS also remain confidential among different layers. This is because none of the Bobs can infer the outcomes of the other. As a result, the  corresponding bits which contribute to secrets  also remain unknown.  So, the secrets shared in different layers remain confidential.
The sifted rate is $1$ bit in all the layers as two symbols are generated with equal probability.
 \subsection{Security against various attacks}
 \label{security}
In this section, we discuss the robustness of the protocol against various eavesdropping strategies. Similar to the LSQKD protocol, Eve's strategy would be to identify participants who belong to all the layers. However, in this particular network, there is no participant who belongs to all three layers. Therefore, if Eve wants to retrieve information being shared in all the layers, she may attack two layers together as in LSQKD, and one layer separately. We consider various eavesdropping strategies one by one.\\
 
 \subsubsection{Intercept-resend attack}
 
 Since all Bobs measure only in the computational basis, Eve also measures the intercepted states in the computational basis and sends post-measurement states to respective Bob. With this strategy, Eve gets the shared information. However, her actions will be detected. This is because Bob performs a measurement only in half of the rounds, which are randomly chosen. In the rest of the rounds, he simply sends the received states back to Alice. When Alice performs   projective measurements on the received states, she will not get the states she had prepared, thus revealing Eve's presence. If Eve intercepts the states of only one Bob, she introduces errors with a probability of $0.75$. Therefore, in $l$ such rounds, Eve gets detected with a probability $(1-0.25^l)$. However, to know the secret being shared, Eve needs to intercept the states of all Bobs. If she does so, her actions get detected with a probability $(1-0.25^{xl})$ for $l$ such rounds. Here, the symbol $x$ represents the number of Bobs whose states are intercepted.   

Similar to the LSQKD protocol, the entangle-and-measure attack has the same effect as that of the intercept-and-resend attack.
Therefore, we now focus on a more general, two-way entangling attack.
\subsubsection{Two-way entangling attack}
  The LSQSS protocol is designed for a network of three layers-- layer $L_1$ (Bob$_1$ and Bob$_2$), layer $L_2$ (Bob$_{3-5}$) and layer $L_3$ (Bob$_{1-5}$). In the following, we consider a particular case in which Eve wants to know the secret being shared in layer $L_1$.\\
  
  \noindent{\textit{Attack in layer $L_1$:}} In layer $L_1$, a secret is generated with outcomes of both the Bobs, therefore, Eve needs to attack subsystems of both the Bobs.  
  
  Let $U$ and $V$ be the two transformations with which Eve interacts her ancillae with the subsystems traveling to and from Bob$_1$ and Bob$_2$ respectively. Since Eve is interested only in the information shared in layer $L_1$, we ignore the states of the other Bobs for the rest of the discussion.

  The state sent to Bob$_1$ and Bob$_2$ is,
  \begin{equation}
      \ket{\phi}=\frac{1}{4}\Big(\ket{0}+\ket{1}+\ket{2}+\ket{3}\Big)^{\otimes 2}=\frac{1}{4}\sum_{i,j=0}^3
\ket{i}\ket{j}  \end{equation}

  Since Eve is interested in knowing the secret in layer $L_1$, she interacts her ancillae with subsystems of Bob$_1$ and Bob$_2$ with unitary $U$. The combined state of Bob$_1$, Bob$_2$, and Eve after interaction can be expressed as:
  \begin{equation}
      \ket{\phi}\ket{0}_{E_1}\ket{0}_{E_2}\xrightarrow[]{U}\frac{1}{4}\sum_{i,j=0}^3\ket{ij}\ket{E_i^{(1)}}\ket{E_j^{(2)}},
  \end{equation}
 where $\ket{E^{(1)}_{i}},~\ket{E^{(2)}_{j}}$ are states of Eve's ancillae after interaction with Bob$_1$ and Bob$_2$'s subsystems respectively.\\
 Similarly, the action of Eve's interaction $V$ on the states traveling from Bob$_1$ and Bob$_2$ to Alice can be expressed as:
 \begin{equation}
     \ket{kl}\ket{0}_{F_1}\ket{0}_{F_2}\xrightarrow[]{V}\sum_{i,j=0}^3\ket{ij}\ket{F^{(1)}_{ki}}\ket{F^{(2)}_{lj}},
 \end{equation}
where  $\ket{0}_{F_1}$ and $\ket{0}_{F_2}$ are states of Eve's ancillae before interaction. The states $\ket{F^{(1)}_{ki}}$ and  $\ket{F^{(2)}_{lj}}$ are unnormalized and non-orthogonal states of Eve's ancillae after interaction with subsystems of Bob$_1$ and Bob$_2$ respectively.

Since each Bob performs CTRL and Reflect operations randomly, we discuss these cases one by one as follows:\\

\noindent{\textit{(a) Bob$_1$ and Bob$_2$ perform CTRL operations:}} The rounds in which Bob$_1$ and Bob$_2$ perform CTRL operations contribute to the generation of secrets. Let the post-measurement states of Bob$_1$ and Bob$_2$ be $\ket{i}$ and $\ket{j}$ respectively. 
If Alice also measures in the computational basis, Eve's interventions get detected when all three participants (Alice and both Bobs) compare a subset of rounds on a classical channel. This is because by doing so, Alice gets the same result with a probability of $p=\norm{\ket{F^{(1)}_{ii}}\ket{F^{(2)}_{jj}}}^2$. So, Eve's interventions get detected with probability $(1-p)$.\\

\noindent{\textit{(b) Bob$_1$ and Bob$_2$ perform Reflect operations:}}
When both Bobs perform Reflect operations, Eve's two interactions act sequentially on the state $\ket{\phi}$. The combined state of Bob$_1$, Bob$_2$, and Eve after these interactions can be expressed as:
\begin{align}
   \ket{\phi'}=\frac{1}{4}\sum_{i,j,k,l=0}^3\ket{kl}\ket{E^{(1)}_i}\ket{E^{(2)}_j} \ket{F^{(1)}_{ik}}\ket{F^{(2)}_{jl}}
\end{align}
If Alice performs a projective measurement $\ket{\phi}\bra{\phi}$, she gets the correct result with probability,
$$p=\norm{\frac{1}{16}\sum_{i,j,k,l=0}^3\ket{E^{(1)}_i}\ket{E^{(2)}_j}\ket{F^{(1)}_{ik}}\ket{F^{(2)}_{jl}}}^2.$$ Thus, tampering of Eve gets detected with a probability $(1-p)$.

\noindent{\textit{(c) Bob$_1$ and Bob$_2$ performs CTRL and Reflect operation:}}
When Bob$_1$ performs CTRL operation and Alice measures in the computational basis, Eve introduces errors in the statistics of Bob$_1$ and Alice. Suppose that Bob$_1$ gets post-measurement state $\ket{i} $ upon performing a CTRL operation. Alice does not get the same state as sent by Bob$_1$ with a finite probability $p=\sum_{j\neq i}\norm{\ket{F^{(1)}_{ij}}}^2$ which is an indicator of Eve's tampering.

A similar analysis holds when Bob$_1$ performs Reflect operation and Bob$_2$ performs the CTRL operation.

\subsubsection{Participant attack}
 It may so happen that a participant may try to provide false information, however, the secret provider,  $\textit{viz.,}$ Alice can detect this attack. To do so, Alice, in some rounds, can send specific states in computational basis. This is because Bobs measurement does not disturb these states. Since these states are sent randomly, so, if Bob provides false information about his outcomes, he gets detected as Alice knows the state deterministically. In this way, such an attack can be detected.

Since LSQSS employs multi-dimensional separable states, it can be implemented with weak coherent pulses (WCP). However, in such scenarios, the protocol becomes vulnerable to implementation-based attacks. In the following, we discuss how the protocol can be made robust against these attacks.

\subsubsection{Photon-number-splitting attack}\label{photon}
Since WCPs are not single-photon sources, Eve can exploit this feature to gain information. Eve may block all single-photon pulses and keep one photon from multi-photon pulses with herself uninterrupted. She can measure this photon after basis reconciliation step. In this manner, Eve may gain full information being shared without introducing any errors. This attack can be detected as already proposed in \cite{wang2005beating} by employing an additional WCP with a lower mean photon number. Since Eve cannot distinguish the two, her action of blocking single-photon pulses gets detected when Alice and the respective Bob compare the signal received for two WCPs. In this manner, the photon-number-splitting attack of Eve can be detected.

\subsubsection{Trojan horse attack}\label{Trojan}
 Similar to the last case, Eve may employ attacks such as delay photon or invisible photon attacks. The delay photon attack can be detected by using randomly counting the number of photons for a subset of rounds as discussed in  \cite{qin2019quantum}. In an invisible photon attack, Eve may want to gain information about Bob's choice by sending  photons of different frequencies.  Such attacks can be prevented as explained in  \cite{qin2019quantum} using a filter of the specified frequency.

This concludes our discussion of layered semi-quantum secret sharing protocol. We now move on to a protocol designed for layered networks consisting of both honest and dishonest participants.


\section{Integrated layered semi--quantum key distribution and secret sharing (ILSKSS)}\label{LSQSS+KD}
In the previous sections, we show that by employing multidimensional states, keys/ secrets can be distributed in multiple layers of a network having only one QP in one go. However, these multidimensional states can be utilized to realize more general tasks, for example, for the distribution of both keys as well as secrets in a single network simultaneously. In the following, we show that the task of key distribution and secret sharing in a single network can be integrated and realized by a suitable choice of multidimensional entangled states. This particular protocol acts as a template for more general situations in which different tasks need to be integrated as per the requirements.\\

\noindent{\textbf{\textit{Network:}}} A network of four participants, Bob$_1$, Bob$_2$, Bob$_3$, and Bob$_4$ and two layers $L_1$ and L$_2$. Layer $L_1$ consists of Bob$_1$ and Bob$_2$ and layer $L_2$ consists of all four Bobs. A pictorial representation of the network is given in figure (\ref{fig:iks}).\\

\begin{figure}[h!]
\centering
\includegraphics[width=0.4\textwidth]{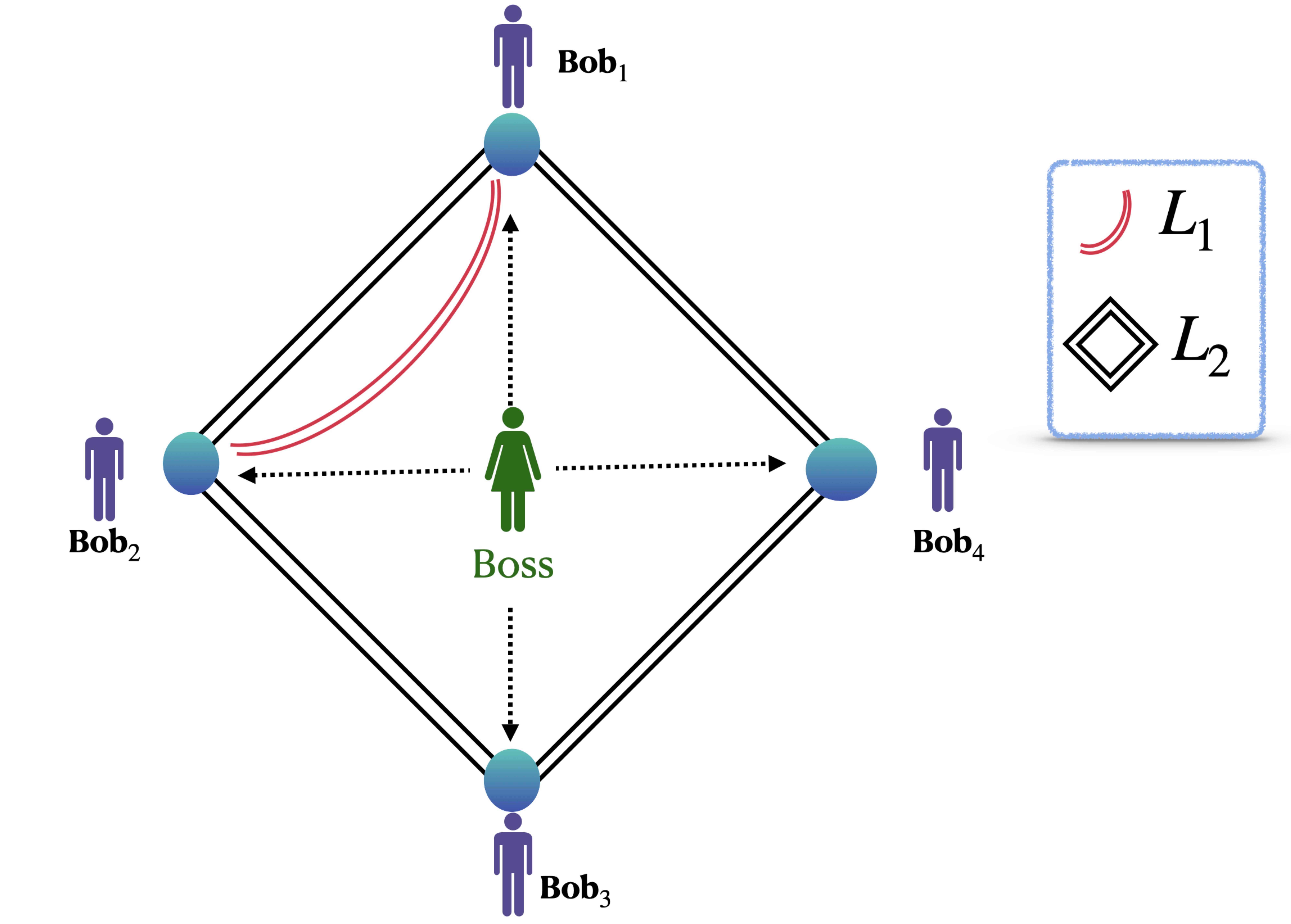}
\caption{Pictorial representation of a network having four participants and two layers. Alice is the external QP who distributes a secret and a key in layers $L_1$ and $L_2$ respectively. Different layers are shown in the inset.}
\label{fig:iks}
\end{figure}

\noindent{\textbf{\textit{Aim:}}} Simultanaeous distribution of a secret in layer $L_1$ and a key in layer $L_2$.\\

\noindent{\textbf{\textit{Resources:}}} The multidimensional entangled state,
 \begin{align}
     |\chi\rangle&=\frac{1}{4}\Big\{\Big(|00\rangle+|02\rangle+|20\rangle+|22\rangle \nonumber\\     &+|11\rangle+|13\rangle+|31\rangle+|33\rangle\Big)|00\rangle\nonumber\\     &+\Big(|01\rangle+|10\rangle+|03\rangle+|30\rangle\nonumber\\
&+|12\rangle+|21\rangle+|23\rangle+|32\rangle\Big)|11\rangle\Big\},
 \end{align}
 acts as a resource. 
The rationale for this choice of state has been given in Appendix (A3).
 \begin{center}
     \textbf{The protocol}
 \end{center}
 The steps of the protocol are as follows:
 \begin{enumerate}
     \item Alice prepares the state $|\chi\rangle$ and sends the $i^{{\rm th}}$ subsystem to Bob$_i$ ($i\in\{1,2,3,4\}$). 
     \item Each Bob performs CTRL or Reflect operations randomly.\\ Please recall that \textit{CTRL} operation corresponds to measurement in the computational basis and sending the post-measurement state to Alice. On the other hand, the  \textit{Reflect} operation corresponds to simply sending the received state to Alice.
     \item Alice, after receiving the subsystems,  either performs measurements on each subsystem in the computational basis or performs a projective measurement, $\Pi_\chi=|\chi\rangle\langle\chi |$.
\item Steps $(1-3)$ constitute one round. They are repeated for a sufficiently large number of rounds.
     \item Afterwards, each Bob reveals the rounds in which CTRL operations are performed and Alice reveals the rounds in which she has measured in the computational basis.
     \item Alice analyses the data of those rounds, in which none of the Bobs have performed CTRL operations, to check for the presence of an eavesdropper.
         \item In the absence of any eavesdropping, the rounds, in which all the Bobs of a given layer perform CTRL operations and Alice measures in the computational basis, generate  symbols for the same layer.
   \end{enumerate}

 \subsection{Key and secret generation rule}
 Similar to the previous cases, all the Bobs rewrite their measurement outcomes in the binary representation. Let the outcomes of Bob$_1$, Bob$_2$, Bob$_3$, and Bob$_4$ be $b_1,~b_2,~b_3$, and $b_4$ respectively. In the binary representation, these are expressed as\footnote{ The outcomes $b_3,~b_4$ of Bob$_3$, and Bob$_4$ are already in the binary form. }:
 \begin{align}
   b_1=2b_1^{(1)}+b_1^{(0)},\quad b_2=2b_2^{(1)}+b_2^{(0)}.
 \end{align}
 A secret in layer $L_1$ is defined as $s_1\equiv b_1^{(1)}\oplus b_2^{(1)}$.
 Similarly, a key in layer $L_2$ is  given by $k\equiv b_1^{(0)}\oplus b_2^{(0)}=b_3=b_4$. \\

\noindent{\textbf{\textit{Confidentiality of keys:}}}
 The secret shared between Bob$_1$ and Bob$_2$ is completely confidential from Bob$_3$ and Bob$_4$. This is because Bob$_3$ and Bob$_4$ cannot obtain any information about the measurement outcomes of Bob$_1$ and Bob$_2$. 

 \subsection{Security against various attacks}
 \label{security}
Similar to the previous two protocols, we discuss the robustness of the protocol against various eavesdropping strategies. We start with the intercept-resend attack.\\

\subsubsection{Intercept-resend attack} 
In the protocol, both keys and secrets are shared in the two layers. If Eve wants to retrieve the key being shared in the layer $L_2$ through the intercept-resend attack, it suffices for her to attack the subsystem of either Bob$_3$ or Bob$_4$. However, Eve's this interception gets detected when all Bobs chose Reflect operation and Alice performs the projective measurement $\ket{\chi}\bra{\chi}$ with a probability of $0.5$. So, in $l$ rounds, her actions get detected with a probability of $(1-0.5^l)$.

Similarly, if Eve wishes to know both key and secret being shared in layer $L_1$, she needs to attack subsystems of both Bob$_1$ and Bob$_2$. Similar to the previous case, her interventions introduce errors with a probability of $1-\frac{1}{16}$. So, in $l$ such rounds, Eve gets detected with a probability of $(1-\frac{1}{16^l})$.
 
If Eve employs a one-way entangle-and-measure attack, as for the previous two protocols, Eve can be detected with the same probabilities as in intercept-resend attacks.

\subsubsection{Two-way entangling attack} In the following, we show that like LSQKD and LSQSS, the ILSKSS protocol is robust against two-way entangling attacks. Since ILSKSS shares both keys and secret, we show that Eve cannot gain information about either without introducing errors.\\

\noindent{\textit{\textbf{Attack in layer $L_2$:}}} Layer $L_2$ consists of four participants. Eve may attempt to attack subsystems being shared with Bob$_1$ and Bob$_2$ or either of Bob$_3$ or Bob$_4$. Since Bob$_1$ and Bob$_2$ retrieve information only after collaboration, Eve has to attack both Bob$_1$ and Bob$_2$. In that case, Eve introduces more errors in contrast to attacking only one participant. Therefore, it is beneficial for Eve to attack either Bob$_3$ or Bob$_4$.

Suppose that Eve employs unitaries $U$ and $V$ to attack subsystems traveling to and from Bob$_3$.
Let the action of $U$ in the computational basis be expressed as:
\begin{equation}
    \ket{i}\ket{0}_E\xrightarrow[]{U}\ket{i}\ket{E_i},
\end{equation}
where $\ket{E_i}$ are states of Eve's ancilla satisfying constraints imposed by unitary $U$.
Since Bob$_3$'s subsystem is correlated to all the participants, after this interaction, the combined state of all the participants and Eve can be expressed as:
\begin{align}
    \ket{\chi}_U=&\frac{1}{\sqrt{2}}\Big\{\ket{\chi_{00}}|00\rangle\ket{E_0}+\ket{\chi_{11}}|11\rangle\ket{E_1}\Big\},
\end{align}
where $\ket{\chi_{00}}$ and $\ket{\chi_{11}}$ represent the states of subsystems of Bob$_1$ and Bob$_2$ and are defined as:
\begin{align}
    \ket{\chi_{00}}\equiv &\frac{1}{2\sqrt{2}}\Big(\ket{00}+\ket{02}+\ket{20}+\ket{22}\nonumber\\
+&\ket{11}+\ket{13}+\ket{31}+\ket{33}\Big)\nonumber\\
    \ket{\chi_{11}}\equiv &\frac{1}{2\sqrt{2}}\Big(\ket{01}+\ket{10}+\ket{03}+\ket{30}\nonumber\\
+&\ket{12}+\ket{21}+\ket{23}+\ket{32}\Big).
\end{align}

The action of Eve's transformation when the subsystems travel from Bob$_3$ to Alice, in the computational basis, can be expressed as:
\begin{align}
    \ket{i}\ket{0}_{F}\xrightarrow[]{V}\sum_{j=0,1}\ket{j}\ket{F_{ij}},
\end{align}
where $\ket{F_{ij}}$ are states of Eve's ancilla after interaction with unitary $V$.
Since Bob$_3$ performs either CTRL or Reflect operation, we analyze the two cases as follows:\\

\noindent{\textit{(a) Bob$_3$ performs CTRL operation:}}
In this case, suppose that Bob$_3$ gets the post-measurement state $\ket{i}$ which he sends to Alice. The rounds in which Alice also measures in the computational basis, due to Eve's interaction, she may not get the same state as sent by Bob$_3$. This discrepancy in the outcomes of the two can be used to check for eavesdropping. That is to say, if Bob$_3$ sends a state $\ket{i}$, Alice gets the same state only with the probability $p=\norm{\ket{F_{ii}}}^2$. Thus, Eve gets detected with a probability $(1-p)$.\\

\noindent{\textit{(b) Bob$_3$ performs Reflect operation:}} When Bob$_3$ performs Reflect operation, Eve's two interactions act sequentially, so in the rounds in which all participants perform Reflect operation, the combined state of all participants and Eve can be expressed as:
\begin{align}    \ket{\chi}\ket{0}_E\ket{0}_{F}\xrightarrow[]{VU}&\frac{1}{\sqrt{2}}\Big\{\ket{\chi_{00}}\Big(\sum_{j=0,1}\ket{j}\ket{0}\ket{E_0}\ket{F_{0j}}\Big)\nonumber\\
&+\ket{\chi_{11}}\Big(\sum_{j=0,1}\ket{j}\ket{1}\ket{E_1}\ket{F_{1j}}\Big)\Big\}.
\end{align}
If Alice measures in the computational basis, she gets the correct result with probability $p=\frac{1}{2}\big(\norm{\ket{E_0}\ket{F_{00}}}^2+\norm{\ket{E_1}\ket{F_{11}}}^2\big)$. Thus, Eve gets detected with a probability $(1-p)$.

However, if Alice performs the projective measurement on $\ket{\chi}\bra{\chi}$, she gets the correct result with a probability $p=\norm{\frac{1}{2}\Big(\ket{E_0}\ket{F_{00}}+\ket{E_1}\ket{F_{11}}\Big)}^2$. Thus, Eve gets detected with a probability $(1-p)$.\\

\noindent{\textbf{\textit{Simultaneous attack in layers $L_1$ and $L_2$:}}} 
In the case, when Eve wants to know both the key and secret being shared in the two layers, it suffices for her to attack subsystems of 
Bob$_1$ and Bob$_2$ as both are participants in the two layers. Please note the difference between the simultaneous attack in the LSQKD and this protocol. Since in ILSKSS, secret is being shared, so Eve cannot gain information without attacking both participants.

Suppose that Eve employs unitaries $U_1$ and $U_2$ to attack subsystem traveling to Bob$_1$ and Bob$_2$ respectively.
The action of these unitaries in the computational basis can be expressed as:
\begin{align}    U_1\ket{i}_{T_1}\ket{0}_{E_1}\equiv\ket{i}\ket{E_{i}},~~~U_2\ket{j}_{T_2}\ket{0}_{E_2}\equiv\ket{j}\ket{E_{j}}
\end{align}
 The combined state of all participants and Eve's ancillae after this interaction can be expressed as:
 \begin{align}
     &\ket{\chi}_{U_1U_2}\equiv \nonumber\\
     &\frac{1}{4}\Big\{\Big(\ket{00}\ket{E_0E_0} +\ket{02}  \ket{E_0E_2}+\ket{20}\ket{E_2E_0}+\ket{22}\ket{E_2E_2}\nonumber\\     &+\ket{11}\ket{E_1E_1}+\ket{13}\ket{E_1E_3}+\ket{31}\ket{E_3E_1}+\ket{33}\ket{E_3E_3}\Big)\ket{00}\nonumber\\
     &+\Big(\ket{01}\ket{E_0E_1} +\ket{10}  \ket{E_1E_0}+\ket{03}\ket{E_0E_3}+\ket{30}\ket{E_3E_0}+\nonumber\\
&\ket{12}\ket{E_1E_2}+\ket{21}\ket{E_2E_1}+\ket{23}\ket{E_2E_3}+\ket{32}\ket{E_3E_2}\Big)\ket{11}\Big\}
     \end{align}
 
 The effect of interaction $V_1$ and $V_2$ in the computational basis on subsystem traveling from Bob$_1$ and Bob$_2$ can be expressed as:
 \begin{align}    V_1\ket{i}_{t_1}\ket{0}_{F_1}\equiv\sum_{j=0}^3\ket{j}\ket{F^{(1)}_{{ij}}},~~~~V_2\ket{i}_{t_2}\ket{0}_{F_2}\equiv\sum_{j=0}^3\ket{j}\ket{F^{(2)}_{{ij}}}
 \end{align}

 \noindent{\textit{(a) Bob$_1$ and Bob$_2$ performs CTRL operation:}} Suppose that when Bob$_1$ and Bob$_2$ perform CTRL operations, they get post-measurement states $\ket{ij}$,  which are sent to Alice. If Alice also measures in the computational basis, she gets the correct result only with a probability $p=\norm{\ket{F^{(1)}_{{ii}}F^{(2)}_{jj}}}^2$. Thus, Alice and Bob can detect Eve's presence by comparing a subset of outcomes with a probability $(1-p)$.\\

 \noindent{\textit{(b) Bob$_1$ and Bob$_2$ perform Reflect operation:}} In this case, Bob$_1$ and Bob$_2$ left the state undisturbed and send it to Alice.
 If Alice performs the projective measurement $\ket{\chi}\bra{\chi}$, she gets the correct result with probability $p=\norm{\frac{1}{16}\sum_{i,j=0}^3\ket{E_iE_j}\ket{F^{(1)}_{ii}F^{(2)}_{jj}}}^2$. So, Eve gets detected with  a probability $(1-p)$.

This concludes the discussion of security of the ILSKSS protocol against various eavesdropping strategies. Given the protocols and their robustness, we now move on to present generalizations of these protocols to arbitrarily  structured layered networks.


\section{Generalisation of protocols to an arbitrary layered structure}\label{generalisation}

We have so far illustrated the protocols with specific numbers of participants and layers. 
In this section, we generalize these protocols to a network having any number of layers and participants. 
Consider a network, labeled by the symbol $\kappa$, in which secret information is to be distributed. The network, $\kappa$, is characterised by the set $\{n,k,\ell_j\}$, where $n$ represents the number of participants, and $k$ represents the total number of layers in $\kappa$. The symbol $\ell_j$ represents the number of layers to which the participant $u_j$ belongs. Since each layer has exactly one key, the same label is employed for both- the key and the layer.

The steps of the procedure to identify the requisite state for a given protocol are described as follows:
\begin{enumerate}
    \item  We start with the reference multiqubit state, say, $|\phi_i\rangle$ that distributes secure information within $i^{\rm th}$ layer. The states $|\phi_i\rangle$ depend on the protocol under consideration. Their tensor product, $\bigotimes_{i=1}^{k}|\phi_i\rangle$ distributes information in the entire network $\kappa$.
    \item For a given participant $u_j$, we arrange all the qubits $\ket{i_1}, \ket{i_2}, \cdots, \ket{i_{{\ell}_j}}$ sequentially. So, the state  $\bigotimes_{i=1}^{k}|\phi_i\rangle$  assumes the following form,
    \begin{align}
        \bigotimes_{i=1}^{k}|\phi_i\rangle = \sum\alpha_{\{u_1\}\{u_2\}\cdots\{u_n\}}|i_1i_2\cdots i_{\ell_1}\rangle_{u_1}\cdots\ket{i_1i_2\cdots i_{\ell_n}}_{u_n}. 
    \end{align}
  Employing binary-to-decimal mapping, 
  \begin{align}
      |m_1m_2\cdots m_{l_j}\rangle \leftrightarrow \Bigg|\sum_{r=1}^{l_j}2^{l_j-r}m_{r}\Bigg\rangle\equiv |m\rangle,\quad m_r\in\{0,1\},
        \label{Mapping}
  \end{align}
     the state $\bigotimes_{i=1}^{k}|\phi_i\rangle$ is reexpressed as a multidimensional state that distributes secure information simultaneously in all the $k$ layers of the network $\kappa$.
\end{enumerate}
   
We now employ this procedure to identify states for each of the three protocols.
\subsection{LSQKD}

Suppose that the reference state, employed to share a key in the $i^{{\rm th}}$ layer is a GHZ state\footnote{If a layer consists of only two participants, the GHZ state reduces to a Bell's state.} given as:
    \begin{align}
        |\Psi_{i}\rangle \equiv \frac{1}{\sqrt{2}}
\left(\bigotimes_j|0\rangle_{u_j}+\bigotimes_j|1\rangle_{u_j}\right),
    \end{align}
    where $u_j$ are the participants belonging to the $i^{{\rm th}}$ layer. The GHZ state is chosen to share a key because it provides a perfect correlation in the computational basis among all the participants.  \\
    Since keys are to be shared in all the $k$ layers, a state that shares keys in a structure $\kappa$ is a tensor product of the GHZ states, each GHZ state sharing a key in a particular layer. That is, the state $|\Psi_\kappa\rangle$ can be written as:
    \begin{equation}
        |\Psi_{\kappa}\rangle \equiv \bigotimes_{i=1}^{k}|\Psi_i\rangle.
    \end{equation}
    Following step 2 of the procedure and  employing the mapping given in equation (\ref{Mapping}), the state $|\Psi_\kappa\rangle$ maps to an $n$--partite multi-dimensional state having layered entanglement. This is the state which is employed to share keys simultaneously in all the layers of a structure $\kappa$. Note that each participant $u_j$ in a network is in possession of $2^{\ell_j}$--dimensional qudit state where $\ell_j$ is the number of layers to which he belongs.
  
 For further clarification, we have illustrated this procedure in Appendix (A1) for a network of three participants in which keys are to be shared in the two layers. 

Similarly, the procedures can be laid down to identify the requisite $n$-partite states that can be employed to implement the rest of the two protocols, {\it viz.}, LSQSS and ILSKSS for any arbitrary structure $\kappa$. In what follows, we describe them one by one.

\subsection{LSQSS}
An LSQSS protocol differs from an LSQKD in that it requires collaboration among all the participants of a given layer to generate a key. This constraint of collaboration is crucial in the choice of a state that shares a secret in a particular layer.

Let, a reference multiqubit state,
\begin{equation}
    |\phi_i\rangle \equiv 
\bigotimes_j\frac{1}{\sqrt{2}}\Big(|0\rangle_{u_j}+|1\rangle_{u_j}\Big),
\end{equation}
be employed to share a secret in the 
$i^{{\rm th}}$ layer of the structure $\kappa$. The symbol $u_j$ ranges over the participants belonging to the $i^{{\rm th}}$ layer. The state is so chosen that none of the participants, based on his outcome, can infer the outcomes of the rest of the participants. Thus, all the participants have to collaborate with one another to share a secret.

Since secrets are to be shared in all the $k$ layers of the network, a state that shares secrets in a structure $\kappa$ is a tensor product of all such states, $\{|\phi_i\rangle\}$. That is, the state $|\Phi_\kappa\rangle$ can be written as:
    \begin{equation}
        |\Phi_{\kappa}\rangle \equiv \bigotimes_{i=1}^{k}|\phi_i\rangle.
    \end{equation}
   As before, following the step 2 of the procedure and employing the mapping given in equation (\ref{Mapping}), the requisite state for the protocol may be identified.

In this way, the state $|\Phi_{\kappa}\rangle$ maps to an $n$-partite multi-dimensional state having a requisite structure that simultaneously shares secrets in a structure $\kappa$. Similar to LSQKD, each participant $u_j$ in a network is in the possession of $2^{\ell_j}$--dimensional qudit state where $\ell_j$ is the number of layers to which he belongs. An illustration of the procedure to identify a state that implements the task of secret sharing in the three layers of a network having five participants is given in Appendix (A2).

\subsection{ILSKSS}
An ILSKSS protocol is a hybrid of both LSQKD and LSQSS protocols. In this protocol, only some of the participants are dishonest, so collaboration is required only among these participants. In this protocol, three kinds of layers are possible--(i) a layer consisting of only the honest participants, (ii) a layer consisting of only the dishonest participants, and (iii) a layer in which some of the participants are dishonest and the rest are honest. The first and the second kind of layers are equivalent to key sharing and secret sharing, whereas the third one requires sharing of a key such that collaboration is required only among the dishonest participants.

  \noindent We identify a reference state for each of the three layers as follows:
    \begin{enumerate}
    \item For each layer $p\in \kappa$ in which all the participants are honest, the same reference state $|\Psi_p\rangle$, as employed in LSQKD can be used, i.e.,
 \begin{align*}
        |\Psi_p\rangle \equiv \frac{1}{\sqrt{2}}
\left(\bigotimes_j|0\rangle_{u_j}+\bigotimes_j|1\rangle_{u_j}\right),
    \end{align*}
     where $j$ ranges over all the honest participants belonging to the layer `$p$'.
        \item  For each layer $q\in \kappa$ in which all the participants are dishonest, the same state, as employed in LSQSS can be used, i.e., the state $|\phi_q\rangle$ is used for secret sharing,
    \begin{align*}
        |\phi_q\rangle \equiv
\bigotimes_t\frac{1}{\sqrt{2}}\Big(|0\rangle_{u_t}+|1\rangle_{u_t}\Big),
    \end{align*}
where $t$ ranges over all the dishonest participants belonging to the layer `$q$'.

    \item For each layer $r\in \kappa$ in which some of the participants are dishonest, collaboration is required only among the dishonest participants. The condition of collaboration leads to the following choice of a resource state,
     \begin{align*}
        |\alpha_{r}\rangle &\equiv \frac{1}{\sqrt{2}}
\Bigg\{\Bigg(\bigotimes_t|+\rangle_{u_t}+\bigotimes_t|-\rangle_{u_t}\Bigg)\bigotimes_j|0\rangle_{u_j}\nonumber\\
&+\Bigg(\bigotimes_t|+\rangle_{u_t}-\bigotimes_t|-\rangle_{u_t}\Bigg)\bigotimes_j|1\rangle_{u_j}\Bigg\},
    \end{align*}
    where,
    \begin{align*}
        |\pm\rangle_{u_t} = \dfrac{1}{\sqrt{2}}\big(|0\rangle_{u_t}\pm |1\rangle_{u_t}\big),
   \end{align*}
    and the index $t$ and $j$ range over the dishonest and honest participants  in the layer `$r$' respectively. The state $|\alpha_r\rangle$ is so chosen that none of the dishonest participants can infer the outcomes of each other as well as that of honest participants. Collaboration among all the dishonest participants is required to infer the outcomes of honest participants, and hence the key.
   \end{enumerate}
 Therefore, the combined state for the structure $\kappa$ is, 
              $$|\chi_{\kappa}\rangle \equiv \bigotimes_{p}|\Psi_p\rangle\bigotimes_{q}|\phi_q\rangle\bigotimes_{r}|\alpha_r\rangle,$$ where the symbols $p$ and $q$ ranges over all the layers consisting of only honest and dishonest participants respectively, and the symbol $r$ ranges over all the layers in which some of the participants are dishonest.
              
              As before, following  step 2 of the procedure and employing the mapping given in equation (\ref{Mapping}), the state $|\chi_{\kappa}\rangle$ maps to an $n$--partite state having the requisite multidimensional layered entangled structure, where $n$ is the number of participants. This state shares secrets and keys in various layers of a structure $\kappa$ in a network simultaneously. For better appreciation, we have provided an illustration that implements the task of sharing a secret in a layer with two participants and sharing a key in another layer with all four participants in the appendix (A3).

We have succinctly shown  the reference multiqubit states (up to normalization) for different protocols in the table (\ref{Reference}).
\begin{center}
\begin{table}[ht!]
\resizebox{7cm}{2.4cm}{
\begin{tabular}{|c| c| c|} 
\hline
\multirow{2}{*}{Protocol} &\multirow{2}{*}{ Nature of participants} & \multirow{2}{4cm}{Reference multiqubit states in$~~~~~$ different layers}  \\
&&\\
\hline
\multirow{2}{*}{LSQKD} & \multirow{2}{*}{Honest }& \multirow{2}{*}{$|\Psi_{i}\rangle \equiv 
\bigotimes_j |0\rangle_{u_j}+\bigotimes_j |1\rangle_{u_j}$}   \\
&&\\
\hline
\multirow{2}{*}{LSQSS} &\multirow{2}{*} {Dishonest }& \multirow{2}{*}{$|\phi_i\rangle \equiv 
\bigotimes_j(|0\rangle_{u_j}+|1\rangle_{u_j})$} \\
&&\\
\hline
\multirow{2}{*}{} & \multirow{2}{*}{Honest } & \multirow{2}{*}{$|\Psi_p\rangle \equiv 
\bigotimes_j|0\rangle_{u_j}+\bigotimes_j|1\rangle_{u_j}$}\\
&&\\
\multirow{5}{*}{ILSKSS}& \multirow{2}{*}{Dishonest} & \multirow{2}{*}{$|\phi_q\rangle \equiv
\bigotimes_t\Big(|0\rangle_{u_t}+|1\rangle_{u_t}\Big)$ }\\ 
&&\\
&  Both   & $|\alpha_{r}\rangle \equiv 
\Big(\bigotimes_t|+\rangle_{u_t}+\bigotimes_t|-\rangle_{u_t}\Big)\bigotimes_j|0\rangle_{u_j}$\\
& & ~~~+$\Big(\bigotimes_t|+\rangle_{u_t}-\bigotimes_t|-\rangle_{u_t}\Big)\bigotimes_j|1\rangle_{u_j}$\\
& & $t$ and $j$ run over\\
& & dishonest and honest participants\\
\hline
\end{tabular}
}
    \caption{Reference multiqubit states for various protocols}
    \label{Reference}
\end{table}
\end{center}

 The combined state to implement ILSKSS in  the structure $\kappa$  is,  $|\chi_{\kappa}\rangle \equiv \bigotimes_{p}|\Psi_p\rangle\bigotimes_{q}|\phi_q\rangle\bigotimes_{r}|\alpha_r\rangle$.

\subsection{Key generation rule}\label{ksrule}
To obtain keys/secrets, each participant rewrites his outcomes in the binary representation. Let $b_i$ be the outcome of the participant Bob$_i$ of a network. In binary representation\footnote{Each participant writes binary representation of his outcome upto the places which is equal to the number of layers he belongs to.}, it is expressed as
$b_i=\sum_{m=0}^{l_j-1}2^mb_i^{(m)}$.
The symbols $b_i^{(m)}$ are used to generate keys/secrets in $(\ell_i-m)^{\rm th}$ layer, where $\ell_i$ is the total number of layers to which Bob$_i$ belongs\footnote{If Alice is also the part of, say $\ell_j$ layers , she also writes her outcome in the binary representation, $a=\sum_{m=0}^{l_j-1}2^m a^{(m)}$. Then, the symbols $a^{(m)}$ are used to generate keys/secrets in $(k-m)^{\rm th}$ layer. }. For all the honest participants, belonging to, say $p^{\rm th}$ layer,  the symbols  $b_i^{(\ell_i-p)}$ correspond to the key symbols.  For all the dishonest participants, belonging to the same $p^{\rm th}$ layer, the key symbol corresponds to $\oplus_{t} b_t^{(\ell_t-p)}$, where $t$ ranges over all the dishonest participants in the $p^{\rm th}$ layer and the symbol $\oplus$ represents the operation of addition modulo 2.  Please note that the keys/secrets shared in different layers remain confidential as in illustrative protocols.

The robustness of these protocols can be shown by a straightforward extension of the robustness analysis for illustrative protocols. 
The difference lies only in the number of participants and layers of a network. For each participant in a given layer, Eve needs to employ two ancillae - one for the forward path (state travel from Alice to respective Bob) and the other for the backward path (state travels back to Alice). Thanks to the random choice of operations by each participant in each round, the action of the eavesdropper introduces errors and hence her presence gets detected. That is, Eve would not be able to gain any information without introducing errors.


 \section{Conclusion}\label{conclusion}
 In summary, this work exploits the potential offered by multidimensional states for proposing semi--quantum communication protocols over a network. We have employed layered entanglement in all the protocols, barring LSQSS which can be implemented with multidimensional product states. We have presented three illustrative protocols and provided a procedure for their generalizations to layered networks of arbitrary structures. We have shown the robustness of these protocols against various eavesdropping attacks. The extension of these protocols to  device-independent scenarios constitutes an interesting study.

 Integration of communication protocols constitutes a rich structure and opens up avenues for further study of quantum communication. This work opens up possibilities of integration of various protocols, being quantum or semi--quantum, as per the requirement of a network. 

\section*{
DATA AVAILABILITY STATEMENT}
Data sharing is not applicable to this article as no datasets were generated or analyzed during the current study.
\section*{Conflict of interest}
The authors declare no conflict of interest.
\section*{Acknowledgements}
Rajni acknowledges financial support from the University Grants Commission in the initial stages of the work. Sooryansh acknowledges support from the Council for Scientific and Industrial Research India (File no.: 09/086(1278)/2017-EMR-I) in the initial stages of the work. We thank the referees for their valuable comments which have helped in bringing clarity to the manuscript.

\section*{Appendices}
\section*{A1. Equivalence of tripartite layered entangled state with tensor product of EPR and three--party GHZ state}\label{secA1}
In section (\ref{LSQKD}), we have proposed the LSQKD protocol for the distribution of keys in the two layers of a three-party network. The protocol employs multi-dimensional entangled state. In the following, employing mathematical equivalence between the proposed protocol and two QKD protocols running in parallel, we identify the requisite multi-dimensional entangled state.

To distribute key in a layer with two participants, a maximally entangled Bell state can be employed. Similarly,  a tripartite GHZ state can be used to distribute a key in a network of three participants. 

 We show that by taking these two states as reference states, we identify the requisite state that distributes keys in the two layers simultaneously. We start with the reference states,
\begin{align}
     |\Psi_{1}\rangle &=\frac{1}{\sqrt{2}}(|0_{A'}0_{B'_1}\rangle+|1_{A'}1_{B'_1}\rangle),\quad\nonumber\\
     |\Psi_{2}\rangle &=\frac{1}{\sqrt{2}}(|0_A0_{B_1}0_{B_2}\rangle+|1_A1_{B_1}1_{B_2}\rangle),
 \end{align}
 where the subscripts represent the participants to which qubits belong. The primes in the subscripts of the state $|\Psi_1\rangle$ are used to differentiate the subsystems from that of the state $|\Psi_2\rangle$.

 The combined state that distributes keys in the two layers is,
 \begin{align}\label{eq:equivalence}
     &|\beta\rangle=|\Psi_{1}\rangle|\Psi_{2}\rangle=\frac{1}{2}\Big(|0_{A'}0_{B'_1}\rangle+|1_{A'}1_{B'_1}\rangle\Big)\nonumber\\
&~~~~~~~~~~~~~~~~~~~~~~~~~~~~~~~~~~~~~~\Big(|0_A0_{B_1}0_{B_2}\rangle+|1_A1_{B_1}1_{B_2}\rangle\Big)\nonumber\\
     &=\frac{1}{2}\Big(|0_{A'}0_{B'_1}0_A0_{B_1}0_{B_2}\rangle+|0_{A'}0_{B'_1}1_A1_{B_1}1_{B_2}\rangle\nonumber\\
     &+|1_{A'}1_{B'_1}0_A0_{B_1}0_{B_2}\rangle+|1_{A'}1_{B'_1}1_A1_{B_1}1_{B_2}\rangle\Big)\nonumber\\
     &=\frac{1}{2}\Big(|0_{A'}0_A0_{B'_1}0_{B_1}0_{B_2}\rangle+|0_{A'}1_A0_{B'_1}1_{B_1}1_{B_2}\rangle\nonumber\\
&+|1_{A'}0_A1_{B'_1}0_{B_1}0_{B_2}\rangle+|1_{A'}1_A1_{B'_1}1_{B_1}1_{B_2}\rangle\Big).
 \end{align}
 We employ the mapping defined in equation (\ref{Mapping}) of section (\ref{generalisation}), which for two qubits system translates to, 
 \begin{equation}
     |00\rangle\rightarrow |0\rangle,\quad |01\rangle\rightarrow\ket{1},\quad\ket{10}\rightarrow\ket{2},\quad\ket{11}\rightarrow\ket{3}.
 \end{equation}
 
Under this mapping, the multiqubit state $\ket{\beta}$ gets mapped to the following  equivalent multiqudit state,
 \begin{align}\label{eq:equi}
     \ket{\beta}&\rightarrow \frac{1}{2}\Big(\ket{000}+\ket{111}+\ket{220}+\ket{331}\Big)\equiv \ket{\Psi},
 \end{align}
  which is the requisite state employed in the LSQKD protocol. 
\section*{A2. Identification of the requisite state for LSQSS}\label{secA2}
To identify the requisite state that shares secrets in all the three layers in a network of five participants, we start with three states, each sharing a secret in the respective layer.
The three states are:
\begin{align}
    \ket{\phi_1}&= \bigotimes_{i=1,2}\frac{1}{\sqrt{2}}(\ket{0}+\ket{1})_{B_i} ,\quad\quad\ket{\phi_2}= \bigotimes_{i=3}^5\frac{1}{\sqrt{2}}(\ket{0}+\ket{1})_{B_i}\nonumber\\
    \ket{\phi_3}&= \bigotimes_{i=1}^5\frac{1}{\sqrt{2}}(\ket{0}+\ket{1})_{B'_i},
\end{align}
where the subscripts represent the participant to which qubits belong. The primes in the subscripts of the state $\ket{\phi_3}$ are used to differentiate its subsystems from that of the states $\ket{\phi_1}$ and $\ket{\phi_2}$. The combined state that distributes secrets in the three layers is,
\begin{align}
   \ket{\gamma}&\equiv \ket{\phi_1}\ket{\phi_2}\ket{\phi_3} \nonumber\\
   &= \bigotimes_{i=1,2}\frac{1}{\sqrt{2}}(\ket{0}+\ket{1})_{B_i} \bigotimes_{i=3}^5\frac{1}{\sqrt{2}}(\ket{0}+\ket{1})_{B_i}\bigotimes_{i=1}^5\frac{1}{\sqrt{2}}(\ket{0}+\ket{1})_{B'_i}\nonumber
\end{align}
Clubbing the subsystems belonging to same participants together, and employing the mapping  $ \ket{00}\rightarrow \ket{0},~ \ket{01}\rightarrow\ket{1},~\ket{10}\rightarrow\ket{2},~\ket{11}\rightarrow\ket{3}$, the state $\ket{\gamma}$ maps to,
\begin{align}
   \ket{\gamma}&\rightarrow \bigotimes_{i=1}^5\frac{1}{2}(\ket{0}+\ket{1}+\ket{2}+\ket{3})_{B_i}  \nonumber\\
   &=\Big(\frac{1}{2}(\ket{0}+\ket{1}+\ket{2}+\ket{3})\Big)^{\otimes 5},
\end{align}
which is same as the state $\ket{\Phi}$ in equation (\ref{eq:SQSS}).
\section*{A3. Identification of the requisite state for ILSKSS}\label{secA3}
In section (\ref{LSQSS+KD}), a protocol has been proposed that distributes a secret and  a key in two layers of a network having four participants. The four participants are named Bob$_1$, Bob$_2$, Bob$_3$, and Bob$_4$. The first layer consists of two dishonest participants named Bob$_1$, and Bob$_2$. The second layer consists of all the four participants. To identify a state that implements the task, we start with reference states,
\begin{align}
    &\ket{\phi_1}\equiv \ket{++}_{B'_1B'_2}=\frac{1}{2}(\ket{00}+\ket{01}+\ket{10}+\ket{11})_{B'_1B'_2},\nonumber\\&\ket{\alpha}\equiv\frac{1}{2}\Big\{\Big(\ket{00}+\ket{11}\Big)\ket{00}+\Big(\ket{01}+\ket{10}\Big)\ket{11}\Big\}_{B_1B_2B_3B_4}\nonumber
\end{align}

Thus, a state that implements the task is,
\begin{align}
    \ket{\zeta}&\equiv\ket{\phi_1}\ket{\alpha}\nonumber\\
    & =\frac{1}{4}\Big(\ket{00}+\ket{01}+\ket{10}+\ket{11}\Big)_{B'_1B'_2}\nonumber\\
    &~~~~\Big((\ket{00}+\ket{11})\ket{00}+(\ket{01}+\ket{10})\ket{11}\Big)_{B_1B_2B_3B_4}
\end{align}
Clubbing the subsystem belonging to the same participant and employing the mapping,
$ \ket{00}\rightarrow \ket{0},\quad \ket{01}\rightarrow\ket{1},\quad\ket{10}\rightarrow\ket{2},\quad\ket{11}\rightarrow\ket{3}$, the state $\ket{\zeta}$ maps to,
\begin{align}
    \ket{\zeta}\rightarrow &\frac{1}{4}\Bigg\{\Big(\ket{00}+\ket{02}+\ket{20}+\ket{22}+\ket{11}+\ket{13}+\ket{31}+\ket{33}\Big)\ket{00}\nonumber\\
     &+\Big(\ket{01}+\ket{10}+\ket{03}+\ket{30}+\ket{12}+\ket{21}+\ket{23}+\ket{32}\Big)\ket{11}\Bigg\},
\end{align}
which is the same state as $\ket{\chi}$.

\end{document}